\documentclass{article}
\usepackage{amsmath,amsfonts,amsthm,amssymb}
\usepackage[mathscr]{eucal}

\newtheorem{thm}{Theorem}
\newtheorem{cor}[thm]{Corollary}
\newtheorem{lem}[thm]{Lemma}
\newtheorem{prop}[thm]{Proposition}
\newcommand{\R}{\mathbb{R}}

\newcommand{\N}{\mathbb{N}}

\newcommand{\eq}[1]{(\ref{#1})}
\newcommand{\supp}{\operatorname{supp}}
\newcommand{\Sym}{\operatorname{Sym}}
\newcommand{\E}{\mathbb{E}}
\newcommand{\Var}{\operatorname{Var}}

\begin{document}

\title
{The Extended Variational Principle for Mean-Field, Classical Spin
Systems}

\author{\vspace{5pt} E.~Kritchevski$^{1}$ and S.~Starr$^{2}$\\
\vspace{-2pt} {$^{1}$Department of Mathematics and Statistics}\\
\vspace{-2pt} {McGill University}\\
\vspace{-2pt} {805 Sherbrooke Street West}\\
\vspace{5pt}  {Montreal, Qu\'eb\'ec, H3A 2K6, Canada.}\\
\vspace{-2pt} {$^{2}$Department of Mathematics}\\
\vspace{-2pt} {University of California, Los Angeles}\\
\vspace{-2pt} {Box 951555}\\
\vspace{5pt}  {Los Angeles, CA, 90095-1555, USA.}\\
    {Email: \texttt{ekritc@math.mcgill.edu},
    \texttt{sstarr@math.ucla.edu}}}

\maketitle

\abstract {The purpose of this article is to obtain a better
understanding of the extended variational principle (EVP). The EVP
is a formula for the thermodynamic pressure of a statistical
mechanical system as a limit of a sequence of minimization
problems. It was developed for disordered mean-field spin systems,
spin systems where the underlying Hamiltonian is itself random,
and whose distribution is permutation invariant. We present the
EVP in the simpler setting of classical mean-field spin systems,
where the Hamiltonian is non-random and symmetric. The EVP
essentially solves these models. We compare the EVP with another
method for mean-field spin systems: the self-consistent mean-field
equations. The two approaches lead to dual convex optimization
problems. This is a new connection, and it permits a
generalization of the EVP.}

\tableofcontents

%%%%%%%%%%%%%%%%%%%%%%%%%%%%%%%%%%%%%%%%%%%%%%%%%%%%%%%%%
%%%%%%%%%%%%%%%%%%%%%%%%%%%%%%%%%%%%%%%%%%%%%%%%%%%%%%%%%
%%%%%%%%%%%%%%%%%%%%%%%%%%%%%%%%%%%%%%%%%%%%%%%%%%%%%%%%%

\section{Introduction}

The extended variational principle (EVP) was introduced in
\cite{AS2}, by Aizenman, Sims, and one of the present authors. It
was applied to a mean-field disordered spin system, known as the
Sherrington-Kirkpatrick spin glass. This is an Ising spin system,
whose underlying Hamiltonian is random, such that the joint
distribution of the coupling constants is permutation symmetric.
The purpose of the EVP there was to give a variational formulation
of the pressure, different than the usual Gibbs variational
principle (GVP). For spin glasses, it seems that the GVP does not
yield a useful characterization of the pressure because of the
complicated dependence of that formula on the random coupling
constants.

The EVP was used to re-derive upper bounds on the quenched
pressure originally proved by Guerra in \cite{Guerra}. Also, the
proof in \cite{AS2} helps to unify that bound with the earlier
proof of existence of the quenched pressure by Guerra and
Toninelli \cite{GuerraToninelli}. Moreover, the approach of
\cite{AS2} introduced the new concept of ``random overlap
structures" of which, Ruelle's random probability cascade (RPC)
\cite{Ruelle} seems to give distinguished examples, having certain
invariance properties. On the other hand, the sequence of
variational formulas that comprise the EVP are still difficult to
work with. For example, the Euler-Lagrange equations were not
derived.

Shortly after the preprint for \cite{AS2}, Talagrand announced a
proof of the most interesting problem related to the
Sherrington-Kirkpatrick model, namely ``Parisi's ansatz". (C.f.,
Talagrand's paper \cite{Talagrand} and his book
\cite{TalagrandBook}.) This does not diminish interest in the EVP
and its relation to spin-glasses. There is hope that the new
insight which could be gained by finding a proof of Parisi's
ansatz based on the EVP and random overlap structures would lead
to more general results.

Since the Euler-Lagrange equations are so hard to determine for
mean-field spin glasses, it seems like a good idea to consider
mean-field classical spin systems, where the situation is easier.
These are Ising-type spin systems (and generalized versions) where
the Hamiltonian is non-random, and permutation symmetric. It turns
out that for such systems, not only can the Euler-Lagrange
equations be derived, they can be essentially ``solved".

For the spin systems just described, there is another method of
solution, called the ``self-consistent mean-field equations". It
consists of solving an implicit, self-consistency equation for a
$1$-body measure. One way to quickly derive the implicit formula
is to write down the GVP. The GVP requires one to optimize a
certain function over the set of all permutation-invariant
$N$-body measures. But instead one optimizes just over the
restricted manifold of $N$-body product measures. The
Euler-Lagrange equation for the GVP on this restricted set gives
the self-consistent mean-field equation. Often one cannot
explicitly solve this 1-body problem, but the mere fact that it
reduces an $N$-body problem to a 1-body problem justifies calling
this a ``solution". The solution obtained by the EVP is similar in
that it also reduces the $N$-body problem to a $1$-body problem.

In the course of our research, we were led\footnote{We are very
grateful to B.~Nachtergaele for bringing this paper to our
attention.} to the beautiful and concise paper of Fannes, Spohn
and Verbeure which treats mean-field quantum spin systems and
gives a rigorous justification of the the self-consistent
mean-field equations. By specialization, their results also apply
to classical spin systems. In the classical case\footnote{In the
quantum case, their method uses a generalization of de Finetti's
theorem by St{\o}rmer \cite{Stormer}, and an alternative to the
Gibbs formulation suitable for quantum spin systems by two of
those authors \cite{FV1,FV2}.}, their method uses the Gibbs
variational formula, combined with de Finetti's theorem. We will
call this the Gibbs, de Finetti principle (GdFP), henceforth.

The de Finetti theorem says the following. Consider a countable
number of spins, indexed by sites of $\N$, say. Then the measure
on $\Omega^{\N}$ is called ``exchangeable" if it is permutation
invariant, for permutations of the arguments which fix all but a
finite number of them. The limit Gibbs measures all have this
property by virtue of the underlying symmetry of the Hamiltonians.
The de Finetti theorem says that the most general exchangeable
measure is a mixture of i.i.d., product measures on the spins
indexed by $\N$. With further work, one can restrict attention to
the extreme measures, which are i.i.d., product measures (so that
the mixture is trivial).

One of the goals of our paper is to compare the EVP  to the GdFP.
Before describing the comparison, let us mention two other useful
approaches to solving mean-field spin systems, which we will not
discuss in this paper. One approach is the ``coherent states
approach", which is useful for quantum spin systems. This was
worked out by Lieb in \cite{Lieb} for the large-spin limit of the
Heisenberg model, and was also applied to the Dicke Maser model by
Hepp and Lieb in \cite{HeppLieb2}. In fact, it seems to be Hepp
and Lieb's work on the Dicke Maser model which motivated Fannes,
Spohn and Verbeure. The other approach uses large deviation
estimates. A good reference is  Ellis and Newman's paper on the
Curie-Weiss model, \cite{EllisNewman}. An advantage of \cite{FSV}
is its generality.

Coming back to the comparison of the EVP and the GdFP, let us say
that both give the same information, when they work. This leads
one to expect that there may be a more direct link between the two
approaches. Indeed there is. It is simplest to see in the $2$-body
case, when the interaction defines a convex bilinear form on
measures. Then the two problems can be viewed as dual optimization
problems in the sense of convex variational analysis. More
precisely, there is a joint ``Lagrangian" which is a
concave-convex function of two variables. Maximizing over the
concave variable gives the nonlinear function which one needs to
minimize in the EVP. Minimizing over the convex variable yields
the nonlinear function which one needs to maximize in the GdFP. So
the fact that both methods lead to the same quantity -- the
thermodynamic pressure -- is a consequence of the fact that the
max-min of the joint Lagrangian equals the min-max.

In case the interaction is continuous and bounded, it is trivial
to see that the min-max and the max-min are equivalent, even in
the non-convex case, and for $n$-body interactions with $n>2$. But
for singular interactions, the equality is nontrivial.
Nevertheless, it is true, and follows from a theorem called the
Kneser, Fan theorem. This theorem is a generalization of the
famous von Neumann minimax theorem. This allows one to generalize
the extended variational principle to some models with singular
interactions (e.g., Coulomb repulsions).

As the reader will see, the EVP is easy to understand and prove,
because it only uses estimates based on convexity and Jensen's
inequality. In comparison, to prove the GdFP one must use
properties of relative entropy, as well as the de Finetti theorem.
Therefore, the latter is more complicated than the former. On the
other hand, the GdFP is more robust.

In conclusion, we would like to make one extrapolation to spin
glasses, which is the following: it would be useful to have an
analogue of de Finetti's theorem, suitable for spin glasses. By
this we mean an intrinsic characterization of the limiting
measures in spin-glasses in terms of an invariance principle with
respect to some stochastic dynamics. (Note that the proof of de
Finetti's theorem \cite{HewittSavage} actually characterizes the
measures on $\Omega^{\N}$ which are invariant under the shift on
$\N$.) So far, there is one result in this direction. It is the
recent, interesting paper by Aizenman and Ruzmaikina, \cite{AR},
which characterizes the 1-level replica-symmetry-breaking RPC's by
an invariance principle called ``quasi-stationarity".

The layout of this paper is as follows. In Section \ref{Sec:defn}
we give the definition of what we mean by mean-field spin system.
In Section \ref{sec:EVP}, we give the main results related to the
EVP. In Section \ref{Sec:optimizers}, we determine the optimizers
of the EVP, and use this to give a simpler formula for the
pressure. We also state a generalization which we prove later, for
singular interactions. In Section \ref{sec:def}, we recall the
main results of the GdFP, as proved in \cite{FSV}  (specialized to
classical spin systems). In Section \ref{sec:comparison} we
construct a joint Lagrangian for the EVP and GdFP. We also prove
the generalization of the EVP from Section \ref{Sec:optimizers}.
In Section \ref{sec:examples}, we give the simplest example.

%%%%%%%%%%%%%%%%%%%%%%%%%%%%%%%%%%%%%%%%%%%%%%%%%%%%%%%%%
%%%%%%%%%%%%%%%%%%%%%%%%%%%%%%%%%%%%%%%%%%%%%%%%%%%%%%%%%
%%%%%%%%%%%%%%%%%%%%%%%%%%%%%%%%%%%%%%%%%%%%%%%%%%%%%%%%%

\section{Mean-Field Spin Systems : Definition} \label{Sec:defn}
%\centerline{\it We will define a mean-field version of a classical
%spin system.}
%
%\smallskip

In this section, we define the notation and set-up for a
``mean-field spin system".
For us a mean-field spin system is defined by
a quadruple $(\Omega,\alpha,n,\phi)$ where: $\Omega$ is a compact
metric space; $\alpha$ is a distinguished Borel probability
measure on $\Omega$ called the {\em a priori} measure; $n$ is a
positive integer determining the number of bodies in the
interaction; and $\phi : \Omega^n \to \R \cup \{+\infty\}$ is the
$n$-body interaction. It is useful that $\Omega$ is a compact
metric space, and that $\alpha$ is a Borel probability measure.
(For example, this means that $\alpha$ is regular.) This is the
level of generality one will find for classical spin systems in
\cite{Israel} and \cite{Simon}.

We denote the set of all Borel probability measures on $\Omega$ by
$\mathcal{M}^+_1(\Omega)$ so that $\alpha \in
\mathcal{M}^+_1(\Omega)$. We will assume that $\phi$ is a Borel
measurable function and that it is bounded from below.
Furthermore, we assume that $\alpha$ and $\phi$ are compatible in
the sense that $\alpha^{\otimes n}(\phi)<\infty$. (Henceforth,
whenever $\mu$ is a measure on a $\sigma$-algebra and $f$ is a
measurable function on the same $\sigma$-algebra, we write
$\mu(f)$  for the integral of $f$ against $\mu$. We also write $f
\mu$ for the [possibly signed] measure such that
$(f\mu)(A)=\mu(f\, \chi_A)$. We use tensor notation to denote
product measures.)

We will assume that $\phi$ is symmetric on $\Omega^n$ with respect
to the natural action of the symmetric group $\mathfrak{S}_n$, as
fits with our intention of studying a {\em mean-field} system. For
each $N \geq n$, we define a Hamiltonian, $H_N : \Omega^N \to \R
\cup \{+\infty\}$. For $x = (x_1,\dots,x_N) \in \Omega^N$,
\begin{equation}
\label{eq:defn:normal:Ham}
  H_N(x)\, :=\, N\, \binom{N}{n}^{-1} \sum_{1\leq i(1)<\dots<i(n)\leq N} \phi(x_{i(1)},\dots,x_{i(n)})\, .
\end{equation}
Note that $H_N$ is symmetric with respect to the natural action of
$\mathfrak{S}_N$. Equivalently, we can think of the underlying
lattice as begin a {\em complete graph}.

For each $N\geq n$, the partition function is the number
$$
  Z(N)\, :=\, \int_{\Omega^N} \exp\left(-H_N(x)\right)\, d\alpha^{\otimes N}(x)
$$
and the finite approximation to the pressure is
$$
  p(N)\, :=\, N^{-1}\, \log\, Z(N)\, .
$$
The {\em thermodynamic pressure} is defined as the limit
$$
  p_*\, :=\, \lim_{N \to \infty} p(N)\, ,
$$
if it exists. We are primarily interested in the thermodynamic
pressure. Later we will recall a well-known result (Theorem
\ref{thm:subadd}) which guarantees that the limit does always
exist.

We have eliminated the inverse-temperature parameter $\beta$, by
absorbing it into the Hamiltonian. It will be fixed and finite for
our entire discussion.

%%%%%%%%%%%%%%%%%%%%%%%%%%%%%%%%%%%%%%%%%%%%%%%%%%%%%%%%%
%%%%%%%%%%%%%%%%%%%%%%%%%%%%%%%%%%%%%%%%%%%%%%%%%%%%%%%%%
%%%%%%%%%%%%%%%%%%%%%%%%%%%%%%%%%%%%%%%%%%%%%%%%%%%%%%%%%

\section{The Extended Variational Principle}
\label{sec:EVP}
\subsection{Setup}

The extended variational principle is a method for calculating the
pressure of a family of Hamiltonians $(\tilde{H}_N\, :\, N)$ which
are close to $(H_N\, :\, N)$. In this section, we will assume that
$\phi$ is a bounded function; i.e., we assume that it is bounded
below, in addition to being bounded above, as in the general
set-up. With this assumption, the new Hamiltonians will be so
close to the old ones that the thermodynamic pressures will be
equal (as we will show).

Let us define a function, $\Phi : \mathcal{M}^{+}_{1}(\Omega) \to
\R$ by
$$
  \Phi(\mu)\, :=\, \mu^{\otimes n}(\phi)\, .
$$
Now, for each $N \in \N_+$, we may define a new Hamiltonian
$\tilde{H}_N : \Omega^N \to \R$ as
\begin{equation}
\label{eq:defn:EVP:Ham}
  \tilde{H}_N(x)\, :=\, N \Phi(\mu_x)\, ,
\end{equation}
where for $x = (x_1,\dots,x_N) \in \Omega^N$,
$$
  \mu_x\, :=\, N^{-1} \sum_{i=1}^N \delta_{x_i}\, .
$$
The measure $\mu_x$ is called the {\em empirical measure} of the
point $x$.

Note that, in the important case that $n=2$, the main difference
between $H_N$ and $\tilde{H}_N$ is the appearance of {\em
self-interaction} terms $\phi(x_i,x_i)$, for $i=1,\dots,N$. One
intuitively expects that these terms make a small contribution,
since there are only $N$ of them, compared to the total number of
terms $N(N-1)/2$. However, if $\phi$ would have an infinite
repulsion, so that $\phi(x_1,x_2)=+\infty$ whenever $x_1=x_2$,
this would lead to a Hamiltonian $\tilde{H}_N \equiv +\infty$,
entirely dominated by the self-interaction terms. This is why we
must assume that $\phi$ is bounded above, as well as below. Of
course, this is a strong requirement, excluding many physically
interesting examples. (In the case $n>2$, the extra terms in
$\tilde{H}_N$ are the natural generalizations of these, where two
or more of the indices coincide.)

We define
$$
  \tilde{Z}(N)\, :=\, \int_{\Omega^N} \exp\left(-\tilde{H}_N(x)\right)\, d\alpha^{\otimes N}(x)\,
  .
$$
We define
\begin{equation*}
    \tilde{p}(N)\, =\, \frac{1}{N}\, \log\,\tilde{Z}(N)\,
    ,
\end{equation*}
and we define $\tilde{p}_*$ as
\begin{equation*}
  \tilde{p}_*\, :=\, \lim_{N \to \infty} \tilde{p}(N)\, ,
\end{equation*}
if it exists.

There is one more \underline{important condition} which we put on
$\phi$. We assume that $\phi$ satisfies the necessary conditions
so that $\Phi:\mathcal{M}^{+}_{1}(\Omega) \to \R$ is either {\em
convex} or {\em concave}. It makes sense to speak of convexity or
concavity of $\Phi$ because $\mathcal{M}^{+}_{1}(\Omega)$ is a
convex set.

\subsubsection{Equivalence of Thermodynamic Pressure} \label{Subsec:equiv}

We will now show the relationship between $p$ and $\tilde{p}$,
under the assumption that $\phi$ is bounded. To begin, we observe
that the energy densities $N^{-1} H_N(x)$ and $N^{-1}
\tilde{H}_N(x)$ are close, in fact
\begin{equation}
\label{eq:eoeH}
    \left|\frac{H_N(x)}{N} - \frac{\tilde{H}_N(x)}{N}\right|\,
    \leq\, \frac{n(n-1)}{N}\, \|\phi\|_{\infty}\, .
\end{equation}
Indeed, this follows because
\begin{equation*}
    \frac{\tilde{H}_N(x)}{N}\, =\,
    \mathbb{E}[\phi(x_{I(1)},\dots,x_{I(n)})]\, ,
\end{equation*}
where the indices $I(1),\dots,I(n)$ are i.i.d.\ random variables,
which are uniform on $\{1,\dots,N\}$, and
\begin{equation*}
    \frac{H_N(x)}{N}\, =\,
    \mathbb{E}[\phi(x_{I(1)},\dots,x_{I(n)})\, |\,
    I(1),\dots,I(n) \textrm{ are distinct}]\, .
\end{equation*}
Therefore,
\begin{equation*}
    \left|\frac{H_N(x)}{N} - \frac{\tilde{H}_N(x)}{N}\right|\,
    \leq\, 2\, \|\phi\|_{\infty}\, \mathbb{P}\{I(1),\dots,I(n) \textrm{ are
    not distinct}\}\, .
\end{equation*}
Then (\ref{eq:eoeH}) follows by bounding the probability,
\begin{equation*}
    \mathbb{P}(\{I(1),\dots,I(n) \textrm{ are
    not distinct}\})\,
    \leq\, \sum_{1\leq j<k\leq n} \mathbb{P}\{I(j)=I(k)\}\, .
%    =\, \frac{1}{N} \binom{n}{2}\, .
\end{equation*}

Now we use an elementary inequality to bound the difference in
$p_N$ and $\tilde{p}_N$, starting from (\ref{eq:eoeH}). But since
we will use the same bound repeatedly hereafter, we will state it
in some some generality.

Suppose $\mathscr{X}$ is a compact metric space, and $\theta \in
\mathcal{M}^+_{1}(\mathscr{X})$ is a Borel probability measure on
$\mathscr{X}$. Define a function, $\Psi$, on the set of Borel
measurable functions $f:\mathscr{X}\to \R \cup \{\pm\infty\}$, as
$$
  \Psi(f)\, :=\, \log\, \theta(e^f)\, .
$$
Then we have the following.
\begin{equation} \label{eq:pressurecompare}
    |\Psi(f) - \Psi(g)|\, \leq\, \|f-g\|_{\infty}\, .
\end{equation}
Indeed, one sees that
\begin{equation*}
    \theta(e^g)\, \leq\, \|e^{g-f}\|\, \theta(e^f)\, ,
\end{equation*}
which proves $[\Psi(g)-\Psi(f)]\, \leq\, \|g-f\|$. The other
inequality follows symmetrically.

Using equation (\ref{eq:pressurecompare}), we see that
\begin{equation*}
    |p(N) - \tilde{p}(N)|\, \leq\, \frac{n(n-1)}{N}\,
    \|\phi\|_{\infty}\, .
\end{equation*}
In particular it implies the following.

\begin{cor} \label{cor:eoe}
Under the assumption that $\|\phi\|_{\infty}<\infty$, the
thermodynamic pressures $p_*$ and $\tilde{p}_*$ either both exist,
or both do not exist, together. In case they both exist, they are
equal.
\end{cor}

\subsection{Results}

%{\it We state the main results of the extended variational
%principle, which are: that the thermodynamic pressure does exist,
%and a formula for calculating it.}
%
%\smallskip

In the bulk of this section we assume that $\Phi$ is convex. In
Subsection \ref{Subsec:Concave}, we will state what changes when
$\Phi$ is concave.

The first main result is the following important fact.
%%%%%%%%%%%%%%%%%%%%%%%%%%%%%%%%%%%%%%%%%%%%%%%
%%%%  EXISTENCE OF THE PRESSURE
%%%%%%%%%%%%%%%%%%%%%%%%%%%%%%%%%%%%%%%%%%%%%%%
\begin{thm}
\label{thm:exists:press} The sequence $(N\, \tilde{p}(N)\, :\, N
\in \N_{+})$ is superadditive. That is, for every pair $N_1,N_2
\in \N_{+}$,
\begin{equation}
\label{eq:superadd}
  (N_1+N_2)\, \tilde{p}(N_1+N_2)\, \geq\,
  N_1\, \tilde{p}(N_1) +
  N_2\, \tilde{p}(N_2)\, .
\end{equation}
Moreover the sequence, $(\tilde{p}(N)\, :\, N\in\N)$ converges in
$\R$.
\end{thm}

\noindent {\bf Remark:} {\it Compare to the main theorem in
\cite{GuerraToninelli}. Also compare to \cite{BCG}.}

\medskip

It is a  well-known fact that for a superadditive sequence
$(X(N)\, :\, N \in \N_+)$, the limit of $N^{-1} X(N)$ exists,
although possibly equal to $+\infty$. (See the origianl by Fekete
\cite{Fekete}, or problem \#98 of P\'olya and Szeg\"o \cite{PS}
[for which there are English translations].) Therefore, the
importance of the second part of the theorem is that the limit is
not $+\infty$.

The second main result is a variational formula for $\tilde{p}_*$.
To set this up, we require some definitions. We first note that,
since $\Omega$ is a compact metric space, the weak topology on
$\mathcal{M}^{+}_{1}(\Omega)$ is compact and metrizable. (C.f.,
\cite{RS} Section IV.4 and \cite{DS} Section V.5.) Thinking of
$\mathcal{M}^{+}_{1}(\Omega)$ as a compact metric space, we define
$\mathcal{M}^{+}_{1}(\mathcal{M}^{+}_{1}(\Omega))$ as the set of
Borel probability measures on it, which is also compact and
metrizable with the weak topology. We also define
$\mathcal{M}^{+}_{f}(\mathcal{M}^{+}_{1}(\Omega))$ to be the set
of all (positive) Borel measures, $\rho$, such that
$$
  0\, <\, \rho(\mathcal{M}^{+}_{1}(\Omega))\, <\, \infty\, .
$$
This is a cone whose base is the Choquet simplex
$\mathcal{M}^{+}_{1}(\mathcal{M}^{+}_{1}(\Omega))$.

The main idea behind the extended variational principle is a
physical notion called the cavity step. Following the prescription
in \cite{AS2}, we will define a sequence of functions, which we
call the cavity field functions. There is a different cavity field
function for each $N \in \N_+$ corresponding to adding $N$ extra
particles to a system, whose size is supposed to be much larger
than $N$.

If the original system is large enough, then instead of
considering a configuration in $\Omega^M$ for some large $M$, we
instead consider a measure in $\mathcal{M}^{+}_{1}(\Omega)$. Note
that because the Hamiltonian is permutation invariant, we only
ever need to consider configurations in $\Omega^M$ modulo
permutations. But using the {\it empirical measures} one can embed
the quotient space $\Omega^M/\mathfrak{S}_M$ into
$\mathcal{M}^{+}_{1}(\Omega)$ for each $M$. Since
$\mathcal{M}^{+}_{1}(\Omega)$ contains each of these finite
configurations spaces, it does make sense to consider a {\em
large} system size by replacing configurations in some $\Omega^M$
by measures in $\mathcal{M}^{+}_{1}(\Omega)$.

It is useful to define $\Phi^{(1)} : \mathcal{M}^{+}_{1}(\Omega)
\times \mathcal{M}^{+}_{1}(\Omega) \to \R$ as
$$
  \Phi^{(1)}(\nu,\mu)\, =\, n \, [\nu^{\otimes n-1} \otimes \mu](\phi)\, .
$$
This is the directional derivative of $\Phi(\nu)$ in the direction
$\mu$. I.e.,
$$
  \frac{d}{dt} \Phi(\nu+t \mu)\, \Big|_{t=0}\,
  =\, \Phi^{(1)}(\nu,\mu)\, .
$$
For each $N \in \N_+$, we define two functions from
$\mathcal{M}^{+}_{f}(\mathcal{M}^{+}_{1}(\Omega))$ to $\R$. These
are
\begin{equation*}
%\label{eq:G1:defn}
  \tilde{G}_N^{(1)}(\rho)\, :=\, N^{-1} \log\,
  \int_{\mathcal{M}^{+}_{1}(\Omega)} \int_{\Omega^N} \exp\left[-N\, \Phi^{(1)}(\nu,\mu_x)\right]\,
  d\alpha^{\otimes N}(x)\, d\rho(\nu)\, ;
\end{equation*}
and
\begin{equation*}
  \tilde{G}_N^{(2)}(\rho)\, :=\, N^{-1} \log\,
  \int_{\mathcal{M}^{+}_{1}(\Omega)} \exp\left(-N\, \left[\Phi^{(1)}(\nu,\nu)
  - \Phi(\nu)\right]\right)\, d\rho(\nu)\, .
\end{equation*}
We define the cavity field function (for addition of $N$ particles
to a large system) as
\begin{equation}
  \tilde{G}_N(\rho)\, :=\, \tilde{G}_N^{(1)}(\rho) - \tilde{G}_N^{(2)}(\rho)\, .
\end{equation}
This function is {\em homogeneous of degree-0}. This means that
\begin{equation*}
    \forall  \rho \in
    \mathcal{M}^{+}_{f}(\mathcal{M}^{+}_{1}(\Omega))\, ,\
    \forall t \in (0,\infty)\quad :\quad
    \tilde{G}_N(\rho)\, =\, \tilde{G}_N(t \rho)\, .
\end{equation*}
This fact is obvious because scaling by $t$ simply adds the same
constant to each of $\tilde{G}_N^{(1)}(\rho)$ and
$\tilde{G}_N^{(2)}(\rho)$, which cancels in the difference.

For every measure $\rho \in
\mathcal{M}^{+}_{f}(\mathcal{M}^{+}_{1}(\Omega))$, there exists a
$t \in (0,\infty)$ such that $t \rho$ is actually a probability
measure. Therefore, we could restrict attention to
$\mathcal{M}^{+}_{1}(\mathcal{M}^{+}_{1}(\Omega))$. But it is
sometimes useful to be free of the constraint that all measures
should be normalized. One easily sees that $\tilde{G}_N$ is
bounded on $\mathcal{M}^{+}_{1}(\mathcal{M}^{+}_{1}(\Omega))$
using equation (\ref{eq:pressurecompare}). Therefore, using
homogeneity, it is bounded on
$\mathcal{M}^{+}_{f}(\mathcal{M}^{+}_{1}(\Omega))$. Moreover,
using the monotone class theorem, and the fact that $\Phi$ is
Borel-measurable, one can check that $\tilde{G}_N$ is
Borel-measurable (c.f., \cite{LL} Section 1.3). If $\phi$ is
continuous, then $\Phi$ is continuous, and it is clear that then
$\tilde{G}_N$ is also continuous.

The main theorem for this section is the following characterization of the pressure.
%%%%%%%%%%%%%%%%%%%%%%%%%%%%%%%%%%%%%%%%
%%%%  EXTENDED VARIATIONAL PRINCIPLE
%%%%%%%%%%%%%%%%%%%%%%%%%%%%%%%%%%%%%%%%
\begin{thm}[EVP]
\label{thm:EVP} For each $N \in \N_+$,
\begin{equation} \label{eq:EVP:easy:ineq}
  \tilde{p}(N)\, \leq\, \inf_{\rho} \tilde{G}_N(\rho)\, ,
\end{equation}
where the infimum is taken over  $\rho \in
\mathcal{M}^{+}_{f}(\mathcal{M}^{+}_{1}(\Omega))$. Moreover,
\begin{equation}
\label{eq:EVP:hard}
  p_*\, =\, \tilde{p}_*\, =\, \lim_{N\to\infty} \inf_{\rho_N} \tilde{G}_N(\rho_N)\, ,
\end{equation}
where, for each $N \in \N$, we infimize over $\rho_N \in
\mathcal{M}^{+}_{f}(\mathcal{M}^{+}_{1}(\Omega))$ separately.
\end{thm}

\noindent {\bf Remark:} {\it Compare to the main theorem in
\cite{AS2}.}

\medskip

We will prove this theorem, as well as Theorem
\ref{thm:exists:press}, in the next section. First we state what
changes if $\Phi$ is concave instead of convex.

%%%%%%%%%%%%%%%%%%%%%%%%%%%%%%%%%%%%%%%%%%%%%%%%%%%%%%%%%
%%%%
%%%%  CHANGES FOR THE CONCAVE CASE
%%%%
%%%%%%%%%%%%%%%%%%%%%%%%%%%%%%%%%%%%%%%%%%%%%%%%%%%%%%%%%
\subsubsection{Changes for the concave case} \label{Subsec:Concave}

In the concave case the sequence of finite approximations to the
pressure is subadditive instead of superadditive, so that the
inequality in (\ref{eq:superadd}), from Theorem
\ref{thm:exists:press}, is reversed. In the extended variational
principle, Theorem \ref{thm:EVP}, the inequality of
(\ref{eq:EVP:easy:ineq}) is reversed, and the infimum is replaced
by the supremum. The identity in (\ref{eq:EVP:hard})  still holds,
but with the infimum replaced by the supremum.

\subsection{Proofs}

All proofs are exactly symmetric between the convex and concave
cases for $\Phi$. So we will only give proofs of the convex case.

\smallskip

\noindent {\bf Proof of Theorem 4.1.}
%{\it Superadditivity follows from convexity of $\Phi$.}
One needs to show $\tilde{Z}(M+N) \geq \tilde{Z}(M)\,
\tilde{Z}(N)$ for every $M,N \in \N_+$. I.e.,
\begin{multline*}
    \int_{\Omega^{M+N}} \exp(-\tilde{H}_{M+N}(z))\, d\alpha^{\otimes
    M+N}(z)\\
    \geq\,
    \int_{\Omega^M\times \Omega^N} \exp(-[\tilde{H}_N(x) + \tilde{H}_M(y)])\,
    d\alpha^{\otimes N}(x)\, d\alpha^{\otimes M}(y)\, .
\end{multline*}
Rewriting $z = (x,y)$, this inequality follows by proving
$$
  \tilde{H}_{M+N}((x,y))\, \leq\, \tilde{H}_N(x) + \tilde{H}_M(y)\, ,
$$
which is equivalent to
\begin{equation}
\label{eq:super:pf2}
  (M+N) \Phi(\mu_{(x,y)})\, \leq\,
  N \Phi(\mu_x) + M \Phi(\mu_y)\, ,
\end{equation}
using the definition \eq{eq:defn:EVP:Ham}. But $\mu_{(x,y)}$ is a
convex combination
\begin{equation*}
  (M+N)\cdot \mu_{(x,y)}\, =\, N\cdot \mu_x + M\cdot \mu_y\, .
\end{equation*}
So \eq{eq:super:pf2} follows from convexity of $\Phi$ and Jensen's
inequality.

%{\it Finiteness follows from an a priori bound.}
It is a well-known fact that for superadditive sequences, $(X(N)\,
:\, N \in \N)$, the limit of $N^{-1} X(N)$ exists, although
possibly equal to $+\infty$. See \cite{Fekete}, or see Lemma
\ref{lem:super:properties}, below. Therefore, $(\tilde{p}(N)\, :\,
N \in \N)$ converges in $\R \cup \{+\infty\}$. But there are
obvious upper bounds which rule out the limit $+\infty$, namely
$\tilde{p}(N)\leq \|\phi\|_{\infty}$. $\blacksquare$

\medskip

The first half Theorem \ref{thm:EVP} is easy to prove. We  only
need to use convexity.

\medskip
\noindent {\bf Proof of Theorem \ref{thm:EVP}, Equation
\eq{eq:EVP:easy:ineq}.}
%{\it Again, this will follow from convexity of $\Phi$.}
It suffices to show that
$$
\tilde{p}(N) + \tilde{G}_N^{(2)}(\rho)\, \leq\,
\tilde{G}_N^{(1)}(\rho)\, ,
$$
for every $\rho \in
\mathcal{M}^{+}_{f}(\mathcal{M}^{+}_{1}(\Omega))$. Direct
calculation yields
\begin{multline*}
    \tilde{p}(N) + \tilde{G}_N^{(2)}(\rho)\\
    =\, N^{-1} \log\, \int_{\mathcal{M}_{+,1}(\Omega)} \int_{\Omega^N}
    \exp\left(-N\,  f(\nu,x)\right)\, d\rho(\nu)\, d\alpha^{\otimes N}(x)\, ,
\end{multline*}
where
$$
  f(\nu,x)\, =\, \Phi(\mu_x) - \Phi(\nu) + \Phi^{(1)}(\nu,\nu)\, .
$$
Similarly,
$$
  \tilde{G}_N^{(1)}(\rho)\,
   =\, N^{-1}\, \log\, \int_{\mathcal{M}_{+,1}(\Omega)} \int_{\Omega^N}
   \exp\left(-N\, \Phi^{(1)}(\nu,\mu_x)\right)\, d\rho(\nu)\, d\alpha^{\otimes N}(x)\, .
$$
Therefore, the inequality holds by showing that
\begin{equation}
\label{convex:consequence}
  \Phi(\mu_x) - \Phi(\nu) - \Phi^{(1)}(\nu,\mu_x-\nu)\, \geq\, 0\, .
\end{equation}
But one easily checks that for $0<t<1$
\begin{equation*}
  \frac{d}{dt} \Phi(t\cdot \mu_x + (1-t)\cdot \nu)\,
  =\, \Phi^{(1)}(t\cdot \mu_x + (1-t)\cdot \nu;\mu_x-\nu)\, .
\end{equation*}
Using this, (\ref{convex:consequence}) is a standard consequence
of convexity of $\Phi$. $\blacksquare$

\medskip
For the proof of the second half of Theorem \ref{thm:EVP}, we will
rely on the following lemma. Although it is well-known that the
limit $N^{-1} X(N)$ exists when $(X(N)\, :\, N \in \N_+)$ is a
superadditive sequence, there is another simple fact which is not
as well-known, but which is essential to the extended variational
principle. This was used, notably, in \cite{AS2}. We repeat the
proof here, for completeness.

\begin{lem}
\label{lem:super:properties} Let $(X(N)\, :\, N \in \N)$ be a
superadditive sequence. Then
$$
  \lim_{N \to \infty} \frac{X(N)}{N}\,
  =\, \lim_{N \to \infty} \liminf_{M \to \infty} \frac{X(M+N) - X(M)}{N}\, .
$$
\end{lem}

\medskip
\noindent {\bf Proof of Lemma \ref{lem:super:properties}.} For
$M,N \in \N_+$, define
$$
  Y(M,N)\, :=\, \frac{X(M+N)-X(M)}{N}
$$
and $Y(N) := \liminf_{M \to \infty} Y(M,N)$. By superadditivity,
$Y(M,N) \geq N^{-1} X(N)$ for all $M \in \N_+$, therefore
\begin{equation}
\label{eq:Y:geq:X}
  Y(N)\, \geq\, N^{-1} X(N)\, .
\end{equation}
Suppose that $k,M,N \in \N_+$ and that $r \geq M$. Then by a
telescoping sum
\begin{equation}
\label{eq:telescop}
  Y(r,kN)\,
  =\, \frac{1}{k}\, \sum_{j=0}^{k-1} Y(r+j N,N)\,
  \geq\, \inf_{M' \geq M}\, Y(M',N)\, .
\end{equation}
Given $M,N \in \N_+$, define $\N_+$-valued functions $k,r :
[M+N,\infty) \to \N$, which are uniquely specified\footnote{In
Matlab language, $k(n) = \operatorname{div}(n-M,N)$ and $r(n) = M
+ \operatorname{mod}(n-M,N)$.} by the requirements $n = k(n) N +
r(n)$ with the remainder in the range $r(n) \in [M,M+N-1]$. Then
one can easily see
$$
  \liminf_{n \to \infty} \frac{X(n)}{n}\, =\, \liminf_{n \to \infty} \frac{X(n) - X(r(n))}{n-r(n)}
  =\, \liminf_{n \to \infty}\, Y(r(n),k(n) N)\, .
$$
Using \eq{eq:telescop}, this implies
\begin{equation*}
  \liminf_{n \to \infty} \frac{X(n)}{n}\,
  =\, \liminf_{k \to \infty} \min_{r \in [M,M+N-1]} Y(r,k N)\,
  \geq\, \inf_{M' \geq M}\, Y(M',N)\, .
\end{equation*}
Taking the monotone limit in $M$, we obtain
$$
  \liminf_{n \to \infty} \frac{X(n)}{n}\, \geq\, Y(N)\, .
$$
Therefore, by \eq{eq:Y:geq:X}
\begin{equation*}
  \liminf_{n \to \infty} \frac{X(n)}{n}\, \geq Y(N)\, \geq\, \frac{X(N)}{N}\, .
\end{equation*}
Taking the limsup in $N$ shows that $(N^{-1} X(N)\, :\, N \in
\N_+)$ converges. Then by the sandwich theorem, $(Y(N)\, :\, N \in
\N_+)$ also converges, to the same limit. $\blacksquare$

\medskip
It seems that much of the physicists' so-called cavity step is
encoded in Lemma \ref{lem:super:properties}. Using it, we now
complete the proof of Theorem \ref{thm:EVP}.

\medskip
\noindent {\bf Proof of Theorem \ref{thm:EVP}, Equation
\eq{eq:EVP:hard}.} By \eq{eq:EVP:easy:ineq},
$$
  \tilde{p}_*\, = \liminf_{N \to \infty} \tilde{p}(N)\, \leq\,
  \liminf_{N \to \infty}\, \inf_{\rho_N} \tilde{G}_N(\rho_N)\, .
$$
Therefore, \eq{eq:EVP:hard} follows if one can
prove
\begin{equation} \label{eq:EVP:pf:desideratum}
  \tilde{p}_*\, \geq\, \limsup_{N \to \infty}\, \inf_{\rho_N} \tilde{G}_N(\rho_N)\, .
\end{equation}
For each $N \in \N$ suppose there were a sequence of measures
$(\rho^{M}_N\, :\, M \in \N_+)$ such that
\begin{equation}
\label{eq:M:limit:G1}
    \lim_{M\to\infty} \left|\tilde{G}^{(1)}_N(\rho_{N}^{M})
    -\frac{M+N}{N}\, \tilde{p}(M+N)\right|\, =\, 0\, ,
\end{equation}
and
\begin{equation}
\label{eq:M:limit:G2}
    \lim_{M\to\infty} \left|\tilde{G}^{(2)}_N(\rho_{N}^{M})
    -\frac{M}{N}\, \tilde{p}(M)\right|\, =\, 0\, .
\end{equation}
Then it would follow
\begin{equation*}
  \lim_{M \to \infty} \left|\tilde{G}_N(\rho_{N}^{M}) -
  \frac{(M+N)}{N}\, \tilde{p}(M+N) - \frac{M}{N}\, \tilde{p}(M)\right|\,
  =\, 0\, .
\end{equation*}
Taking $(\rho^{M}_N\, :\, M \in \N_+)$ as a variational sequence,
this would imply
\begin{equation*}
  \liminf_{M \to \infty}
  \frac{(M+N) \tilde{p}(M+N) - M \tilde{p}(M)}{N}\,
  \geq\, \inf_{\rho_N} \tilde{G}_N(\rho_N)\, .
\end{equation*}
But by Lemma \ref{lem:super:properties}, applied to $X(N) = N\,
\tilde{p}(N)$, this would give \eq{eq:EVP:pf:desideratum}.
Therefore, it only remains to prove \eq{eq:M:limit:G1} and
\eq{eq:M:limit:G2}.

The map $y \mapsto \mu_y$ is a continuous function from $\Omega^M$
to $\mathcal{M}^{+}_{1}(\Omega)$ (with respect to the weak
topology on the target). Therefore, given any Borel measure
$\tilde{\rho} \in \mathcal{M}^{+}_{f}(\Omega^M)$, there is a
unique measure $\rho \in
\mathcal{M}^{+}_{f}(\mathcal{M}^{+}_{1}(\Omega))$, called the
push-forward, such that
\begin{equation*}
  \int_{\mathcal{M}_{+,1}(\Omega)} F(\mu)\, d\rho(\mu)\,
  =\, \int_{\Omega^M} F(\mu_y)\, d\tilde{\rho}(y)\, ,
\end{equation*}
for every continuous function $F:\mathcal{M}^{+}_{1}(\Omega) \to
\R$ (continuous with respect to the weak topology).

Now consider a measure $\tilde{\rho}_{N}^{M} \in
\mathcal{M}^{+}_{f}(\Omega^M)$, absolutely continuous with respect
to $\alpha^{\otimes M}$, such that
\begin{gather*}
  \frac{d\tilde{\rho}_{N}^{M}}{d\alpha^{\otimes M}}(y)\, =\, \exp\left(-\frac{M}{M+N}\, H_M(y)\right)\, .
\end{gather*}
Let $\rho_{N}^{M} \in
\mathcal{M}^{+}_{f}(\mathcal{M}^{+}_{1}(\Omega))$ be the
push-forward of $\tilde{\rho}_N^{M}$. Then one verifies
\begin{align}
\nonumber
  \tilde{G}_N^{(1)}(\rho_{N}^{M})\,
  &=\, N^{-1} \log\, \int_{\Omega^M} \int_{\Omega^N} \exp\left(-N\, \Phi^{(1)}(\mu_y,\mu_x)\right)\,
  d\alpha^{\otimes N}(x)\, d\tilde{\rho}_{N}^{M}(y)\\
\label{feq}
  &=\, N^{-1} \log\, \int_{\Omega^M\times \Omega^N} e^{-f(x,y)}\, d\alpha^{\otimes N}(x)\, d\alpha^{\otimes M}(y)\, ,
\end{align}
where
$$
  f(x,y)\, =\, \frac{M^2}{M+N}\, \Phi(\mu_y) +  N\, \Phi^{(1)}(\mu_y,\mu_x)\, .
$$
One can prove (\ref{eq:M:limit:G1}) for this sequence of measures.

One can write a formula analogous to equation (\ref{feq}) for
$\tilde{p}(M+N)$, namely
\begin{multline*}
    \frac{M+N}{N}\, \tilde{p}(M+N)\\
    =\, N^{-1} \log\, \int_{\Omega^{M+N}} \exp\left(-(M+N)\,
    \Phi(\mu_z)\right)\, d\alpha^{\otimes M+N}(z)\, .
\end{multline*}
Using the decomposition $z = (x,y)$, this yields
\begin{equation}
\label{geq}
  \frac{M+N}{N}\, \tilde{p}(M+N)\,
  =\,  N^{-1}\, \log\, \int_{\Omega^M\times \Omega^N} e^{-g(x,y)}\, d\alpha^{\otimes N}(x)\, d\alpha^{\otimes M}(y)\, ,
\end{equation}
where $g(x,y) = (M+N)\, \Phi(\mu_{(x,y)})$.

Using the formula
\begin{equation*}
    \mu_{(x,y)}\, =\,  \frac{N \mu_x + M \mu_y}{M+N}\, =\, \mu_y +
    \frac{N}{M+N}\, [\mu_x - \mu_y]\, ,
\end{equation*}
one writes
\begin{align*}
  &\hspace{-25pt} \frac{g(x,y) - f(x,y)}{M+N}\\
  \hspace{25pt} &=\,
  \Phi\left(\mu_y + \frac{N}{M+N}\, [\mu_x-\mu_y]\right)
  - \Phi(\mu_y)
  - \frac{N}{M+N}\, \Phi^{(1)}(\mu_y,\mu_x-\mu_y)\\
  &\quad + \frac{N^2}{(M+N)^2}\, \Phi(\mu_y)\, .
\end{align*}
Now, by Taylor's theorem,
\begin{multline*}
    \Phi\left(\mu_y + \frac{N}{M+N}\, [\mu_x-\mu_y]\right)
  - \Phi(\mu_y) - \frac{N}{M+N}\, \Phi^{(1)}(\mu_y,\mu_x-\mu_y)\\
  =\, \frac{N^2}{(M+N)^2}\, \int_0^1 (1-\theta)\,
  \frac{d^2}{dt^2} \Phi\left(\mu_y + t [\mu_x -
  \mu_y]\right)\, \Big|_{t = \frac{N \theta}{M+N}}\, d\theta\,
  ,
\end{multline*}
and one easily calculates
\begin{multline*}
    \frac{d^2}{dt^2}\, \Phi(\nu+t \mu)\\
    =\,  \binom{n}{2}\, \int_{\Omega^{n-2}} \int_{\Omega^2} \phi(x)\,
  d\nu^{\otimes n-2}(x_1,\dots,x_{n-2})\, d\mu^{\otimes 2}(x_{n-1},x_n)\, .
\end{multline*}
Therefore,
\begin{equation*}
  \|f-g\|_{\sup}\, \leq\, \frac{N^2}{M+N}\, \left[1 + \frac{1}{2} \binom{n}{2}\right]\,
  \|\phi\|_\infty\, .
\end{equation*}
By equation (\ref{eq:pressurecompare}), and equations (\ref{feq})
and (\ref{geq}), this means
$$
  \left|\tilde{G}_N^{(1)}(\rho^{M}_{N}) -  \frac{M+N}{N}\, \tilde{p}(M+N)\right|\,
  \leq\, \frac{N}{M+N}\, \left[1+\frac{1}{2} \binom{n}{2}\right]\, \|\phi\|_{\infty}\, .
$$
This certainly does converge to zero as $M \to \infty$, proving
(\ref{eq:M:limit:G1}). The argument for \eq{eq:M:limit:G2} is
similar, and is left to the reader. $\blacksquare$

\section{Optimizers for the EVP}
\label{Sec:optimizers}

\begin{prop}
\label{prop:EVP:extreme} Suppose that $\phi$ is continuous, and
$\Phi$ is convex or concave. Then, for each $N$, the optimum of
$\tilde{G}_N$ is attained. Moreover, the normalized optimizers
(scaled to be probability measures) form a union of faces of the
Choquet simplex
$\mathcal{M}^{+}_{1}(\mathcal{M}^{+}_{1}(\Omega))$.
\end{prop}

\noindent {\bf Proof.} Restrict attention to the case that $\Phi$
is convex, since the concave case is proved symmetrically. Note
that $\tilde{G}_N$ is continuous. Since
$\mathcal{M}^{+}_{1}(\mathcal{M}^{+}_{1}(\Omega))$ is compact, the
minimum is attained. This proves the first part of the
proposition. The second part of the proposition is equivalent to
the following statement: let $\rho$ be any minimizer in
$\mathcal{M}^{+}_{1}(\mathcal{M}^{+}_{1}(\Omega))$ then for each
$\nu \in \supp(\rho)$, the measure $\delta_{\nu}$ is also a
minimizer.

Let $f$ be any Borel measurable function with $0\leq f\leq 1$ such
that $\rho(f)
> 0$. For $0<t<1$, define $\rho_t \in
\mathcal{M}^{+}_{f}(\mathcal{M}^{+}_{1}(\Omega))$ by
\begin{equation*}
    \frac{d\rho_t}{d\rho}(\mu)\, :=\, (1 + t f(\mu))\, .
\end{equation*}
Then, for $i=1,2$
\begin{equation*}
  \tilde{G}_N^{(i)}(\rho_t)\, =\, \log\left[\exp\left(\tilde{G}_N^{(i)}(\rho)\right)
  + t \exp\left(\tilde{G}_N^{(i)}(\tilde{\rho})\right)\right]\, ,
\end{equation*}
where $\tilde{\rho} \in
\mathcal{M}^{+}_{f}(\mathcal{M}^{+}_{1}(\Omega))$ is the measure
$\tilde{\rho} = f\, \rho$, using an obvious notation.

The two functions, $t \mapsto \tilde{G}_N^{(i)}(\rho_t)$, for
$i=1,2$, are obviously differentiable on $(-1,\infty)$. Therefore
by criticality,
\begin{equation*}
    \frac{d}{dt}\left[\tilde{G}_N^{(1)}(\rho_t)
    - \tilde{G}_N^{(2)}(\rho_t)\right]\Big|_{t=0} \,
    =\, 0\, .
\end{equation*}
But careful consideration of this equation yields
\begin{equation*}
    \tilde{G}_N(\tilde{\rho})\, =\, \tilde{G}_N(\rho)\, .
\end{equation*}
So $\tilde{\rho}$ is another optimizer.

Now, for any $\nu \in \supp(\rho)$ consider the sequence of
functions  $f_{\epsilon} = \chi_{D(\nu;\epsilon)}$ (for
$\epsilon>0$) where $\chi$ is the indicator and $D(\nu;\epsilon)$
is the closed ball, with reference to any metric on
$\mathcal{M}^{+}_{1}(\Omega)$ which yields the weak topology.
(Such a metric is guaranteed to exist since $\Omega$ is compact
and hence separable. C.f., \cite{DS}, Section V.5.) Since $\nu \in
\supp(\rho)$, one knows $\rho(f_{\epsilon})>0$ for all
$\epsilon>0$. The family of rescaled measures
$\rho(f_{\epsilon})^{-1}\, f_\epsilon\, \rho$ converge weakly to
$\delta_{\nu}$ in the $\epsilon \downarrow 0$ limit. Using
continuity of $\tilde{G}_N$, that means $\delta_\nu$ is a
minimizer, as claimed. $\blacksquare$

\medskip
When $\rho$ has the simple form $\delta_{\nu}$ for some $\nu \in
\mathcal{M}^{+}_{1}(\Omega)$, all the values
$\tilde{G}_N(\delta_\nu)$ (for $N \in \N_+$) are identical, and
are given by the function $\tilde{g}(\nu)$ written below.
Therefore, the limit in Theorem \ref{thm:EVP}, equation
(\ref{eq:EVP:hard}) is trivial. We state this as the following:

\begin{cor}\label{cor:nonlinearformula}
Define $\tilde{g} : \mathcal{M}^{+}_{1}(\Omega) \to \R$ by
\begin{equation}
\label{gtilde:def}
  \tilde{g}(\nu)\,
  =\, (n-1) \Phi(\nu)
  + \log\, \int_{\Omega} \exp\left(-\Phi^{(1)}(\nu;\delta_x)\right)\, d\alpha(x)\, .
\end{equation}
Suppose that $\phi$ is continuous and $\Phi$ is convex. Then
\begin{equation*}
    p_*\, =\, \tilde{p}_*\, =\, \min_{\nu} \tilde{g}(\nu)\, .
\end{equation*}
(If $\Phi$ is concave instead of convex, the minimum changes
to the maximum.)
\end{cor}

\subsection{Extension for Convex Two-Body Interactions}

Suppose we drop the restriction that $\phi$ is bounded, and only
require that $\phi : \Omega^n \to \R \cup \{+\infty\}$ is Borel
measurable and bounded below, as in Section \ref{Sec:defn}. In
this case $\tilde{p}_N$ may no longer exist (or rather it may
equal $+\infty$, identically) for each $N \in \N_+$. But the
cavity field function $\tilde{G}_N$ is still well-defined and
finite, if we put certain natural restrictions on the measures
$\rho$ which we use. The same is true for its restriction to
extreme points, defined by $\tilde{g}$. It is reasonable to ask if
one can still determine $p_*$ (which may now be inequivalent to
$\tilde{p}_*$) using $\tilde{g}$? At least in some cases the
answer is, ``yes".

\begin{thm} \label{thm:extension} Suppose $n=2$ and $\Phi$ is convex.
For each $C\geq 0$, define
\begin{equation*}
    \mathcal{M}^{+}_1(\Omega,\alpha,C)\, =\, \left\{\nu \in
    \mathcal{M}^{+}_1(\Omega)\, :\, \nu \ll \alpha\
    \textrm{and}\ \left\|\frac{d\nu}{d\alpha}\right\|_{\infty}
    \leq e^{C}\right\}\, .
\end{equation*}
Then
\begin{equation*}\,
    p_*\, =\, \lim_{C \to \infty}\, \inf_{\nu \in \mathcal{M}^{+}_1(\Omega,\alpha,C)}\, \tilde{g}(\nu)\, .
\end{equation*}
\end{thm}

\noindent {\bf Remarks:} {\it 1.\ Restricting to
$\mathcal{M}^{+}_1(\Omega,\alpha,C)$ is a technical necessity. If
we do not put some restrictions on $\nu \in
\mathcal{M}^{+}_1(\Omega)$, then it is possible that the two
summands in (\ref{gtilde:def}) are $+\infty$ and $-\infty$. On the
other hand, because $\Phi(\alpha)<\infty$, both terms are finite
when $\nu \in \mathcal{M}^{+}_1(\Omega,\alpha,C)$ for some $C$. By
taking the $C \to \infty$ limit, at the end, we relax these
restrictions. This is also the condition that we need in Section
\ref{Sec:minimax}, to apply the Kneser, Fan Theorem.

\smallskip \noindent
2.\ Our proof uses convexity of $\Phi$. It does not give the
analogous statement for the case that $\Phi$ is concave.}

\medskip

The proof of this fact will be given at the end of Section
\ref{Sec:minimax}. It can be seen as the motivation for the
following two sections, though they are also interesting on their
own.

%%%%%%%%%%%%%%%%%%%%%%%%%%%%%%%%%%%%%%%%%%%%%%%%%%%%%%%%%
%%%%%%%%%%%%%%%%%%%%%%%%%%%%%%%%%%%%%%%%%%%%%%%%%%%%%%%%%
%
% The Gibbs Variational Formula and de Finetti's Theorem}
%
%%%%%%%%%%%%%%%%%%%%%%%%%%%%%%%%%%%%%%%%%%%%%%%%%%%%%%%%%
%%%%%%%%%%%%%%%%%%%%%%%%%%%%%%%%%%%%%%%%%%%%%%%%%%%%%%%%%

\section{The Gibbs, de Finetti Principle}
\label{sec:def}
\newcommand{\fH}{\mathcal{H}}
\newcommand{\fZ}{\mathcal{Z}}
\newcommand{\fp}{\mathbf{p}}

In this section we will give a pedagogical introduction to the
paper of Fannes, Spohn and Verbeure \cite{FSV}. In fact, while
they considered quantum spin system, which is more general, we
specialize to the classical case. In order to be self-contained,
we will review the specialization of their results.

\subsection{Setup}

In this section we relax the conditions on $\phi$ relative to the
previous section. We only assume the conditions from Section
\ref{Sec:defn}. Namely, we assume that $\phi : \Omega^n \to \R
\cup \{+\infty\}$ is Borel measurable and bounded below. We
suppose that $\alpha^{\otimes n}(\phi)<\infty$ and that $\phi$ is
invariant under the natural action of $\mathfrak{S}_n$.

We will use two important principles, called the Gibbs variational
formula, and de Finetti's theorem. The Gibbs formula gives a
variational formulation for the finite-volume approximations to
the pressure, $(p(N)\, :\, N\geq n)$. The de Finetti theorem is a
representation theorem for all infinite exchangeable probability
measures. When combined, these two principles give a
mathematically rigorous variational formula for the thermodynamic
pressure of a mean-field classical spin system, which the
physicists also use (but usually without referring to the rigorous
justification).

We start by stating the Gibbs variational formula. The first step
is to recall entropy. Given a measure $\rho^{N} \in
\mathcal{M}^{+}_{1}(\Omega^N)$, its relative entropy with respect
to $\alpha^{\otimes N}$ will be denoted as $S_N(\rho^{N})$.
(Usually the relative entropy would be denoted
$S_N(\rho^{N},\alpha^{\otimes N})$, but we suppress
$\alpha^{\otimes N}$.) This is a quantity in $\R \cup
\{-\infty\}$. If $\rho^N$ is absolutely continuous with respect to
$\alpha^{\otimes N}$, then
$$
  S_N(\rho^N)\,
  :=\, \int_{\Omega^N}\, \psi\left(\frac{d\rho^N}{d\alpha^{\otimes N}}(x)\right)\,
  d\alpha^{\otimes N}(x)\, ,
$$
where
\begin{equation*}
    \psi(t)\, :=\,
    \begin{cases} -t\, \log\, t &\, \textrm{if } t \in (0,\infty]\, ,\\
    0 &\, \textrm{if } t=0\, .\end{cases}
\end{equation*}
Even if $\rho^N \ll \alpha^{\otimes N}$, the relative entropy may
equal $-\infty$ depending on the Radon-Nikodym derivative. If
$\rho^N$ is not absolutely continuous with respect to
$\alpha^{\otimes N}$ (i.e., if the singular component has a
positive mass) then $S_N(\rho^N)$ is defined to be $-\infty$.

Henceforth we will call the quantity ``relative entropy with
respect to $\alpha^{\otimes N}$'' just by the term ``entropy''.
The following important properties of the entropy, except for
Property \ref{property:partition}, are proved in the monographs by
Israel and Simon, respectively: \cite{Israel}, Section II.2, and
\cite{Simon}, Section III.4. The best reference for Property
\ref{property:partition} is the seminal paper by Ruelle and
Robinson, \cite{RuelleRobinson}. One can also consult the
monograph by Georgii \cite{Georgii}, Chapter 15 for related
issues.

As a notational point, for $\rho^N \in
\mathcal{M}^{+}_{1}(\Omega^N)$, and $A \subset [1,N]$, we denote
by $\rho^{N} \restriction A$, the measure in
$\mathcal{M}^{+}_{1}(\Omega^{|A|})$, naturally identified as the
marginal of $\mu^N$ on the $\sigma$-subalgebra of Borel measurable
functions on $\Omega^N$ depending only on coordinates of $x$ for
indices in $A$.
%%%%%%%%%%%%%%%%%%%%%%%%%%%%%%%%%%%%%%%%%%%%%%%%%%
%%%%  PROPERTIES OF RELATIVE ENTROPY
%%%%%%%%%%%%%%%%%%%%%%%%%%%%%%%%%%%%%%%%%%%%%%%%%%
\begin{prop}[Properties of Relative Entropy]
\label{prop:Properties:Entropy} The functions, $S_N :
\mathcal{M}^{+}_{1}(\Omega^N) \to \R \cup \{-\infty\}$ (for $N \in
\N$) have the following properties.
\begin{enumerate}
    \item
    \label{property:partition}
    {\bf (Definition through continuous partitions)}
    \begin{equation*}
        S_N(\rho^N)\, =\, \inf_{R \in \N} \inf_{(u_1,\dots,u_R)}
        \sum_{r=1}^n \psi\left(\rho^N(u_r)/\alpha^{\otimes N}(u_r)\right)\, \alpha^{\otimes N}(u_r)\, ,
    \end{equation*}
    where $(u_1,\dots,u_R)$ varies over all continuous partitions of unity
    on $\Omega^N$, such that $\alpha^{\otimes N}(u_r)>0$ for each $r$.
    \item
    \label{property:maximizer}
    {\bf (Non-positivity)} $S_N(\rho^N) \leq 0$ for all $\rho^N \in
    \mathcal{M}^{+}_{1}(\Omega^N)$ and equality holds for $\rho^N
    = \alpha^{\otimes N}$.
    \item
    \label{property:usc}
    {\bf (Upper semicontinuity)}
    The function $S_N : \mathcal{M}^{+}_{1}(\Omega^N)
    \to \R\cup\{-\infty\}$ is upper semicontinuous with respect to the
    topology of weak convergence.
    \item
    \label{property:concav}
    {\bf (Strict concavity)}
    For $\rho^N_1,\rho^N_2\in\mathcal{M}^{+}_{1}(\Omega^N)$ and
    $\theta \in (0,1)$,
    $$
        S_N(\theta \cdot \rho^N_1+(1-\theta)\cdot \rho^N_2)\,
        \geq\, \theta\, S_N(\rho^N_1)+(1-\theta)\, S_N(\rho^N_2)\, .
    $$
    The inequality is strict if $S_N(\rho^N_i)>-\infty$ for both $i=1,2$,
    unless $\rho^N_1=\rho^N_2$.
    \item
    \label{property:almost:convex}
    {\bf (``Almost convexity")}
    For the setting as above,
    \begin{align*}
        S_N(\theta \cdot \rho^N_1+(1-\theta)\cdot \rho^N_2)\,
        &\leq\, \theta\, S_N(\rho^N_1)+(1-\theta)\,
        S_N(\rho^N_2)\\
        &\qquad + \psi(\theta) + \psi(1-\theta)\, .
    \end{align*}
    \item
    \label{property:subadd}
    {\bf (Strong subadditivity)}
    Given subsets $A,B \subset [1,N]$,
    \begin{multline*}
        S_{|A\cup B|}(\rho^{N} \restriction A \cup B) + S_{|A\cap B|}(\rho^{N}
        \restriction A \cap B)\\ \leq\,
        S_{|A|}(\rho^{N} \restriction A) + S_{|B|}(\rho^{N} \restriction B)\, .
    \end{multline*}
\end{enumerate}
\end{prop}
For this to be consistent, we need to define $S_0$. The need
arises when one takes the marginal $(\rho^{N} \restriction A\cap
B)$ and $A \cap B = \emptyset$. One can make sense of this by
defining $\Omega^0 = \{\emptyset\}$ to be the 1-point space,
defining $\alpha^{\otimes 0}$ to be the unique measure in
$\mathcal{M}^{+}_{1}(\Omega^0)$, and defining $\rho^{N}
\restriction \emptyset$ to be that same measure no matter what
$\rho^N \in \mathcal{M}^{+}_1(\Omega^N)$ may be. Then the
appropriate definition  is obviously $S_0(\rho^{N} \restriction
\emptyset) = 0$ for all $\rho^N \in \mathcal{M}^{+}_1(\Omega^N)$.

Using Property \ref{property:partition} there is a stronger
version of Property \ref{property:concav}. Suppose that $\rho =
\sum_{r=1}^R \theta_r\, \delta_{x_r}$ for $x_1,\dots,x_R \in
\Omega^N$ and $\theta_1,\dots,\theta_R\geq 0$ are such that
$\sum_{r=1}^R \theta_r=1$. Then by iterating Property
\ref{property:concav}
\begin{equation*}
    S_N(\rho^N)\, \geq\, \sum_{r=1}^R \theta_r\,
    S_N(\delta_{x_r})\, .
\end{equation*}
This will be particularly useful in the thermodynamic limit, $N
\to \infty$, when combined with Property
\ref{property:almost:convex}. But we would like to generalize to
allow continuous convex combinations (barycentric
decompositions).

\begin{lem}\label{lem:bary}
Let $(\mathcal{W},\Sigma)$ be a measure space with probability
measure $\theta$. Suppose that there is a measurable mapping
(probability kernel) $w \in \mathcal{W} \mapsto \rho^{N}_w \in
\mathcal{M}^{+}_{1}(\Omega)$. Define the barycenter $\rho^N$ such
that
$$
  \rho^N(f)\, =\, \int_{\mathcal{W}} \rho^N_w(f)\, d\theta(w)
$$
for each $f \in \mathcal{C}(\Omega^N)$. Then,
\begin{equation}
\label{stronger:conc}
  S_N(\rho^N)\, \geq\, \int_{\mathcal{W}} S_N(\rho^N_w)\, d\theta(w)\, .
\end{equation}
\end{lem}

\noindent {\bf Proof.} Let $(u_1,\dots,u_R)$ be a continuous
partition of unity on $\Omega^N$, such that $\alpha^{\otimes
N}(u_r)>0$ for each $r$. By concavity,
\begin{align*}
    &\hspace{-50pt}
    \sum_{r=1}^n \psi\left(\rho^N(u_r)/\alpha^{\otimes N}(u_r)\right)\, \alpha^{\otimes
    N}(u_r)\\
    &\geq\, \sum_{r=1}^n  \left(\int_{\mathcal{W}} \psi\left(\rho^N_w(u_r)/\alpha^{\otimes N}(u_r)\right)\,
    d\theta(w)\right)\, \alpha^{\otimes N}(u_r)\\
    &\geq \int_{\mathcal{W}} S_N(\rho^N_w)\, d\theta(w)\, .
\end{align*}
Since this is true for every such partition, equation
(\ref{stronger:conc}) follows. $\blacksquare$

\medskip

The Gibbs function on $\mathcal{M}^{+}_{1}(\Omega^N)$ is defined
as
\begin{equation}
\label{eq:defn:Gibbs:function}
  G_N(\rho^N)\,
  =\, N^{-1}\, \left[S_N(\rho^N) -
  \rho^N(H_N)\right]\, .
\end{equation}
(Let us reiterate that we have absorbed the inverse temperature
$\beta$ into the Hamiltonian.) The Gibbs measure is a measure
$\rho^{N}_{*} \in \mathcal{M}^{+}_{1}(\Omega^N,\Sym)$ such that
$\rho^N_* \ll \alpha^{\otimes N}$ and
\begin{gather*}
  \frac{d\rho^{N}_{*}}{d\alpha^{\otimes N}}(x)\,
  =\, Z(N)^{-1}\, \exp\left(-H_N(x)\right)\, .
\end{gather*}
Note that, since $\alpha^{\otimes n}(\phi)>0$ we know that
$Z(N)>0$, by an elementary application of Jensen's inequality. An
important formula for statistical mechanics is the following.
%%%%%%%%%%%%%%%%%%%%%%%%%%%%%%%%%%%%%%%%%%%%%%%%%%
%%%%  GIBBS, VARIATIONAL FORMULA
%%%%%%%%%%%%%%%%%%%%%%%%%%%%%%%%%%%%%%%%%%%%%%%%%%
\begin{thm}[Gibbs Variational Formula]
\label{thm:GVP} The Gibbs function is strictly concave and upper
semicontinuous (on the set of measures where it is not equal to
$-\infty$). The maximum is attained at  a unique point, which is
the measure $\rho^{N}_{*}$. Moreover, $\displaystyle
G_N(\rho^N_{*})\, =\, p(N)$.
\end{thm}

\noindent {\bf Proof.} Note that
$$
  G_N(\rho^N)\, :=\, S_N(\rho^N;\rho^N_{*}) + p(N)\, ,
$$
where the first term on the right-hand-side is the relative
entropy with respect to $\rho^N_{\beta}$. All of the properties
from Proposition \ref{prop:Properties:Entropy} are also valid for
relative entropy with respect to measures other than
$\alpha^{\otimes N}$ {\em mutatis mutandis}. The theorem is just a
collection of some of these. (The only thing that changes is the
precise statement of strong subadditivity, which is not used in
this theorem anyway.) $\blacksquare$

\medskip

Having stated the Gibbs formula, let us now state de Finetti's
theorem. To set this up, we will need some notation. If $\rho^N
\in \mathcal{M}^{+}_1(\Omega^N)$ is symmetric under the natural
action of $\mathfrak{S}_N$ on $\Omega^N$, then it is called
``exchangeable''. In this case $\rho^{N} \restriction A$ clearly
only depends on the cardinality, say $R = |A|$. As a notational
simplification, when this is the case, we allow ourselves to write
$\rho^{N \restriction R}$ in place of $\rho^{N} \restriction A$.
We will write the set of all exchangeable measures in
$\mathcal{M}^{+}_1(\Omega^N)$ as
$\mathcal{M}^{+}_1(\Omega^N,\Sym)$.

\medskip
\noindent {\bf Definition:} {\it Given a strictly increasing
sequence $(N(k) \in \N_+\, :\, k \in \N_+)$ and a sequence of
measures $\rho^{N(k)} \in
\mathcal{M}^{+}_{1}(\Omega^{N(k)},\Sym)$, we will say that the
sequence converges weakly if, for every $N \in \N_+$, it happens
that the subsequence of marginals $(\rho^{N(k) \restriction N}\,
:\, k\, ,\ N(k)\geq N)$ converges weakly in
$\mathcal{M}^{+}_{1}(\Omega^{N})$.}

\medskip

Because of properties of the marginal, it will be clear that, if
the sequence of measures $\left(\rho^{N(k)}\, :\, k\in\N_+\right)$
converges weakly, then the weak limits $\rho^{\infty \restriction
N} := \lim_{k \to \infty} \rho^{N(k) \restriction N}$ are
consistent with respect to taking further marginals. Therefore,
the measures satisfy the hypotheses of Kolmogorov's extension
theorem. (C.f.\ \cite{Dudley}, Theorem 12.1.2 or \cite{Stroock},
Exercise 3.1.18.) So there is a naturally identified measure
$\rho^\infty \in \mathcal{M}^{+}_{1}\left(\Omega^{\N}\right)$,
which is defined on the smallest $\sigma$-algebra containing all
cylinder sets (depending on finitely many variables). Moreover
$\rho^{\infty}$ is defined just so that the finite-dimensional
marginals are equal to $\rho^{\infty\restriction N}$, justifying
the notation {\em a posteriori}.

A measure $\rho^{\infty} \in \mathcal{M}^{+}_{1}(\Omega^{\N})$ is
called exchangeable if all of its finite marginals $\rho^{\infty
\restriction N}$ are exchangeable. Let
$\mathcal{M}^{+}_1(\Omega^{\N},\Sym)$ be the set of exchangeable
measures in $\mathcal{M}^{+}_1(\Omega^{\N})$. One may define a
topology on the set of exchangeable measures
%$\mathcal{M}^{+}_{1}(\Omega^{\N},\Sym)$
such that a sequence of
measures $\mu_k^{\infty} \in
\mathcal{M}^{+}_{1}(\Omega^{\N},\Sym)$ converges iff
$\mu_k^{\infty \restriction N}$ converges (as $k \to \infty$) for
each $N \in \N_+$. This topology is metrizable and compact. Indeed
it is the weak topology with respect to the compact metrizable
topology on $\Omega^{\N}$ (c.f., \cite{RS}, Theorem IV.5). The de
Finetti theorem completely characterizes the measures in
$\mathcal{M}^{+}_1(\Omega^{\N},\Sym)$.
%%%%%%%%%%%%%%%%%%%%%%%%%%%%%%%%%%%%%%%%%%%%%%%%%%
%%%%  DE FINETTI'S REPRESENTATION
%%%%%%%%%%%%%%%%%%%%%%%%%%%%%%%%%%%%%%%%%%%%%%%%%%
\begin{thm}[de Finetti's Representation]
\label{thm:deF} For every measure $\rho^{\infty} \in
\mathcal{M}^{+}_{1}(\Omega^{\N},\Sym)$, there is a unique $\rho
\in \mathcal{M}^{+}_{1}(\mathcal{M}^{+}_{1}(\Omega))$, such that
\begin{equation}
\label{eq:deF:rep}
    \rho^{\infty \restriction N}\,
    =\, \int_{\mathcal{M}^{+}_{1}(\Omega)} \mu^{\otimes N}\, d\rho(\mu)\,
    ,
\end{equation}
for every $N \in \N_+$.
%Also, for each $\rho \in
%\mathcal{M}^{+}_{1}(\mathcal{M}_{+,1}(\Omega))$, the measures
%$\rho^{*,N} \in \mathcal{M}_{+,1}(\Omega^N)$ defined as in
%\eq{eq:deF:rep} satisfy the conditions of Kolmogorov's theorem,
%and are exchangeable.
\end{thm}

For a general proof of this theorem, see the paper by Hewitt and
Savage, \cite{HewittSavage}. For many connections to interesting
results in probability theory, see the review of Aldous
\cite{Aldous} and references therein.

\subsection{Results}

The first result, analogous to Theorem \ref{thm:exists:press} is
the following,
%%%%%%%%%%%%%%%%%%%%%%%%%%%%%%%%%%%%%%%%%%%%%%%%%%
%%%%  EXISTENCE OF THE PRESSURE
%%%%%%%%%%%%%%%%%%%%%%%%%%%%%%%%%%%%%%%%%%%%%%%%%%
\begin{thm}
\label{thm:subadd} For every $N_1,N_2 \geq n$,
\begin{equation}
\label{eq:subadd}
  (N_1+N_2)\, p(N_1+N_2)\, \leq\,
  N_1\, p(N_1)(\beta) +
  N_2\, p(N_2)(\beta)\, .
\end{equation}
Also, for each $N\geq n$
\begin{equation}
\label{eq:easy:lower:bd}
  p(N)\, \geq\, -\alpha^{\otimes n}(\phi)\, .
\end{equation}
In particular the sequence $(p(N)\, :\, N\geq n)$ converges in
$\R$.
\end{thm}

\medskip

The second main result of this section is the following formula
for the pressure. In Section \ref{Sec:minimax} this will be
compared to Corollary \ref{cor:nonlinearformula} in the convex
case.
%%%%%%%%%%%%%%%%%%%%%%%%%%%%%%%%%%%%%%%%%%%%%%%%%%
%%%%  GIBBS, DE FINETTI VARIATIONAL PRINCIPLE
%%%%%%%%%%%%%%%%%%%%%%%%%%%%%%%%%%%%%%%%%%%%%%%%%%
\begin{thm}[Gibbs, de Finetti Variational Principle]
\label{thm:GdF} Define $g : \mathcal{M}^{+}_{1}(\Omega) \to \R
\cup \{-\infty\}$ by
$$
  g(\mu)\,
  :=\, S_1(\mu) - \Phi(\mu)\, .
$$
Then, for every $N \geq n$,
\begin{equation}
\label{eq:easy:ineq}
  p(N)\, \geq\, \sup_{\mu} g(\mu)\, ,
\end{equation}
where the supremum is taken over all $\mu \in
\mathcal{M}^{+}_{1}(\Omega)$. The function $g$ is upper
semicontinuous, so the maximum is attained. Moreover,
\begin{equation}
\label{eq:Gibbs:deF}
  p_*\, =\, \max_{\mu}\, g(\mu)\, .
\end{equation}
\end{thm}

%\begin{rem}
%This theorem is a specialization of a result of Fannes, Spohn and Verbeure,
%who proved the quantum version \cite{FSV}.
%For completeness we will reproduce their argument, here, somewhat
%expanded.
%The reader who prefers concise proofs should consult Proposition II.6
%in \cite{FSV}.
%\end{rem}

\subsection{Proofs}

\noindent {\bf Proof of Theorem \ref{thm:subadd}.} Suppose
$\rho^{M+N} \in \mathcal{M}^{+}_{1}(\Omega^{M+N},\Sym)$. By the
definition of the sequence of (permutation invariant)
Hamiltonians,
\begin{align*}
  (M+N)^{-1}\, \rho^{M+N}(H_{M+N})\,
  &=\, M^{-1}\, \rho^{M+N\restriction M}(H_M)\\
  &=\, N^{-1}\, \rho^{M+N\restriction N}(H_N)\, .
\end{align*}
This implies that
\begin{equation*}
  \rho^{M+N}(H_{M+N})\,
  =\, \rho^{M+N\restriction M}(H_M) + \rho^{M+N\restriction N}(H_N)\, .
\end{equation*}
By subadditivity of the entropy, which is Proposition
\ref{prop:Properties:Entropy}, Property \ref{property:subadd}
specialized to the case that $A = [1,M]$ and $B=[M+1,N]$, one
knows
\begin{equation*}
  S_{M+N}(\rho^{M+N})\,
  \leq\, S_M(\rho^{M+N\restriction M}) + S_N(\rho^{M+N \restriction N})\, .
\end{equation*}
Therefore,
\begin{equation}
\label{eq:Gibbs:subadd}
  (M+N)\, G_{M+N}(\rho^{M+N})\,
  \leq\, M\, G_M(\rho^{M+N \restriction M}) + N\, G_N(\rho^{M+N \restriction N})\, .
\end{equation}
Now apply this to $\rho^{M+N}_{*} \in
\mathcal{M}^{+}_{1}(\Omega^{M+N},\Sym)$ and use Theorem
\ref{thm:GVP}. The left-hand-side of \eq{eq:Gibbs:subadd} becomes
$(M+N)\, p(M+N)$, and the two terms on the right-hand-side are
bounded above by $M\, p(M)$ and $N\, p(N)$.

The bound \eq{eq:easy:lower:bd} is a variational lower bound
obtained by the trial $\rho^N = \alpha^{\otimes N}$ and Theorem
\ref{thm:GVP}. From it we know that $p(\beta)>-\infty$, which is
important because subadditive sequences generally may have the
limit $-\infty$, but this one does not.
$\blacksquare$

\medskip

Let us prove the easy part of Theorem \ref{thm:GdF}, which only
uses Theorem \ref{thm:GVP}.

\medskip
\noindent {\bf Proof of Theorem \ref{thm:GdF}, Equation
(\ref{eq:easy:ineq}).} Suppose $\mu \in
\mathcal{M}_{+,1}(\Omega)$. Define $\rho^N = \mu^{\otimes N}$.
Observe that $S_N(\rho^N) = N\, S_1(\mu)$ and
$$
  \frac{1}{N}\, \rho^{N}(H_N)\, =\, \frac{1}{n}\, \rho^{N \restriction
n}(H_n)\,
  =\, \mu^{\otimes n}(\phi)\,
  =\, \Phi(\mu)\, .
$$
So $G_N(\rho^N) = g(\mu)$. Then, using Theorem \ref{thm:GVP}, one
obtains $p(N) \geq g(\mu)$ as a variational lower bound. The
equation follows. $\blacksquare$

\medskip
To prove the second half of Theorem \ref{thm:GdF}, we will use the
following important fact. So far we have only used subadditivity
of the pressure, which is a special case of Theorem
\ref{prop:Properties:Entropy}, Property \ref{property:subadd}. The
next result uses strong subadditivity; in fact it is equivalent to
it.
%%%%%%%%%%%%%%%%%%%%%%%%%%%%%%%%%%%%%%%%%%%%%%%%%%
%%%%  MONOTONE ENTROPY DENSITY
%%%%%%%%%%%%%%%%%%%%%%%%%%%%%%%%%%%%%%%%%%%%%%%%%%
\begin{lem}
\label{lem:Monotone:Entropy:Density} Suppose $N \in \N_+$ and
$\rho^N \in \mathcal{M}^{+}_{1}(\Omega^N,\Sym)$. Then,
\begin{equation}
\label{eq:Monotone:Entropy:Density}
  n^{-1}\, S_n(\rho^{N \restriction n})\,
  \geq\, N^{-1}\, S_N(\rho^N)\, ,
\end{equation}
for every $n \in [1,N]$.
\end{lem}

\noindent {\bf Proof.} It is sufficient to prove that,
\begin{equation}
\label{eq:Str-Subadd:to:Monotone}
  S_N(\rho^{N}) - S_{N-1}(\rho^{N \restriction N-1})\,
  \leq\, S_{N-1}(\rho^{N \restriction N-1}) - S_{N-2}(\rho^{N \restriction N-2})\,
  ,
\end{equation}
for every $N>1$. This is because, by iterating this inequality,
one gets
$$
  S_N(\rho^{N}) - S_{N-1}(\rho^{N \restriction N-1})\,
  \leq\, S_{n}(\rho^{N \restriction n}) - S_{n-1}(\rho^{N \restriction n-1})\, ,
$$
for all $n\leq N-1$. Summing these inequalities over $n\in[1,N-1]$
gives a telescoping sum on the right-hand-side. So
$$
  (N-1)\, S_N(\rho^{N}) - (N-1)\, S_{N-1}(\rho^{N \restriction N-1})\,
  \leq\, S_{N-1}(\rho^{N \restriction N-1}) - S_0(\rho^{N \restriction 0})\, .
$$
By rearranging terms, this would prove
\eq{eq:Monotone:Entropy:Density} when $n=N-1$ (recall that
$S_0(\rho^{N \restriction 0}) := 0$). But then by iterating that,
one could reach all $n\leq N-1$.

It remains to prove \eq{eq:Str-Subadd:to:Monotone}. Use
Proposition \ref{prop:Properties:Entropy}, Property
\ref{property:subadd}, with $A = [1,N-1]$ and $B=[2,N]$.
$\blacksquare$

\medskip
One of the most important consequences of de Finetti's theorem,
for us, is the fact that the relative entropy becomes very simple
in the $N \to \infty$ limit for exchangeable measures. In fact it
is affine. This is expressed in the following lemma, which also
uses Lemma \ref{lem:Monotone:Entropy:Density}.

%%%%%%%%%%%%%%%%%%%%%%%%%%%%%%%%%%%%%%%%%%%%%%%%%%%%%%%%%%
%%%%%
%%%%%%%%%%%%%%%%%%%%%%%%%%%%%%%%%%%%%%%%%%%%%%%%%%%%%%%%%%
\begin{lem}[Mean Entropy]
For every $\rho^{\infty} \in
\mathcal{M}^{+}_{1}(\Omega^{\N},\Sym)$, the following limit exists
$$
  s(\rho^{\infty})\, :=\, \lim_{N \to \infty} N^{-1}\, S_N(\rho^{\infty \restriction N})\, .
$$
The function $s : \mathcal{M}^{+}_{1}(\Omega^{\N},\Sym)\to \R \cup
\{-\infty\}$ is affine and upper semicontinuous. More precisely,
\begin{equation*}
    s(\rho^{\infty})\, =\, \int_{\mathcal{M}^{+}_{1}(\Omega)} S_1(\mu)\,
    d\rho(\mu)\, ,
\end{equation*}
where $\rho \in \mathcal{M}^{+}_1(\mathcal{M}^{+}_1(\Omega))$ is
the ``directing measure" corresponding to $\rho^{\infty}$ via de
Finetti's theorem.
\end{lem}

\noindent {\bf Proof.} The existence of the limit
$s(\rho^{\infty})$ can be proved either by subadditivity, (the
specialization of Proposition \ref{prop:Properties:Entropy},
Property \ref{property:subadd}), or by monotonicity of the entropy
density as in Lemma \ref{lem:Monotone:Entropy:Density}. By the
latter, it is clear that $s$ is upper semicontinuous being the
infimum of upper semicontinuous functions. Also, $s$ is concave by
Proposition \ref{prop:Properties:Entropy}, Property
\ref{property:concav}. Moreover, one can deduce that $s$ is convex
by using Proposition \ref{prop:Properties:Entropy}, Property
\ref{property:almost:convex}, and noting that for the mean entropy
one divides each $S_N$ by $N$, and takes the limit as $N \to
\infty$ (so that the error terms in ``almost convexity" converge
to $0$ uniformly). Therefore, $s$ is affine. Using these
properties and Lemma \ref{lem:bary}, one can prove that
\begin{equation*}
    s(\rho^{\infty})\, =\, \int_{\mathcal{M}^{+}_{1}(\Omega)} s(\delta_\mu)\,
    d\rho(\mu)\, .
\end{equation*}
(Actually, Lemma \ref{lem:bary} only proves that the integral
representation is a lower bound for $s$. But using convexity and
upper semicontinuity, one can easily prove $s(\rho^{\infty})\leq
\max_{\mu \in \supp(\rho)} s(\delta_{\mu})$. By taking the correct
partition, one can then use this to obtain the appropriate
opposite inequality.) But when $\rho = \delta_{\mu}$, one has
$\rho^{\infty\restriction N} = \mu^{\otimes N}$ for all $N$, and
as already noted $S_N(\mu^{\otimes N}) = N\, S_1(\mu)$. So
$s(\delta_{\mu}) = S_1(\mu)$. $\blacksquare$

\medskip
\noindent {\bf Proof of Theorem \ref{thm:GdF}, Equation
\eq{eq:Gibbs:deF}.}  Let $\left(\rho^{N(k)}_*\, :\, k \in
\N_+\right)$ be any weakly convergent subsequence of the Gibbs
measures (which exists because the set of all such sequences is
compact with respect to the topology of weak convergence), and let
$\rho_{*}^{\infty} \in \mathcal{M}^{+}_1(\Omega^{\N},\Sym)$ be the
limit.

Fix $N \geq n$. By Lemma \ref{lem:Monotone:Entropy:Density},
\begin{equation*}
  p(N(k))\,
  =\, G_{N(k)}\left(\rho^{N(k)}_{*}\right)\,
  \leq\, G_N\left(\rho^{N(k) \restriction N}_{*}\right)\, ,
\end{equation*}
for all $k$ such that $N(k)\geq N$. Using Theorem \ref{thm:GVP},
$G_N$ is upper semicontinuous. Therefore,
\begin{equation*}
    \limsup_{k \to \infty}\, p(N(k))\,
    \leq\, G_N(\rho^{\infty \restriction N}_{*})\, .
\end{equation*}
On the other hand, $p(N)$ converges to $p_*$ by Theorem
\ref{thm:subadd}. So
\begin{equation*}
    p_*\, \leq\, G_N(\rho^{\infty\restriction N}_{*})\, .
\end{equation*}
Since this inequality is true for every $N \geq n$, it is also
true that
\begin{equation}
\label{eq:liminf}
    p_*\, \leq\, \inf_{N \in \N_+}
    G_N(\rho^{\infty\restriction N}_*)\, .
\end{equation}
Define another affine, upper semicontinuous function $G_{\infty} :
\mathcal{M}^{+}_{1}(\Omega^{\N}) \to \R \cup \{-\infty\}$ by
$$
  G_\infty(\rho^{\infty})\, =\, s(\rho^{\infty}) - \rho^{\infty \restriction n}(\phi)\, .
$$
Using Lemma \ref{lem:Monotone:Entropy:Density}, one can conclude
that $G_N(\rho^{\infty \restriction N})$ is a decreasing sequence,
converging to $G_{\infty}(\rho^{\infty})$. Then, using this and
(\ref{eq:liminf}),
$$
  p_*\, \leq\, G_{\infty}(\rho^{\infty}_{*})\, .
$$

This is true for each limit point, and there is at least one.
Therefore,
$$
  p_*\, \leq\, \sup_{\rho^{\infty} \in \mathcal{M}^{+}_1(\Omega^{\N},\Sym)} G_{\infty}(\rho^{\infty})\,
  .
$$
Since $\mathcal{M}^{+}_1(\Omega^{\N},\Sym)$ is compact and convex,
and $G_{\infty}$ is a convex (in fact affine) and upper
semicontinuous function, the maximum is achieved, and it is
achieved at an extreme point. By Theorem \ref{thm:deF}, the
extreme points are of the form $\rho^{\infty} = \mu^{\otimes \N}$
for some $\mu \in \mathcal{M}^{+}_{1}(\Omega)$. In other words,
the measure $\rho \in
\mathcal{M}^{+}_{1}(\mathcal{M}^{+}_{1}(\Omega))$ defined via de
Finetti's theorem is $\delta_\mu$ for some $\mu \in
\mathcal{M}^{+}_{1}(\Omega)$ if $\rho^{\infty}$ is an extreme
point of $\mathcal{M}^{+}_1(\Omega^{\N},\Sym)$. In this case, one
can explicitly calculate $G_{\infty}(\rho^{\infty})$. It is
$g(\mu)$. This also proves that $g$ is upper semicontinuous
because it is the restriction of $G_{\infty}$, and that function
is upper semicontinuous. $\blacksquare$

%%%%%%%%%%%%%%%%%%%%%%%%%%%%%%%%%%%%%%%%%%%%%%%%%%%%%%%%%
%%%%%%%%%%%%%%%%%%%%%%%%%%%%%%%%%%%%%%%%%%%%%%%%%%%%%%%%%
%%%%%%%%%%%%%%%%%%%%%%%%%%%%%%%%%%%%%%%%%%%%%%%%%%%%%%%%%

\section{Minimax Theorem and a Joint Lagrangian}
\label{sec:comparison} \label{Sec:minimax}

\subsection{Setup}
Recall that, under the hypothesis that $\|\phi\|_\infty<\infty$,
\begin{equation*}
  p_*\, =\, \tilde{p}_*\, ,
\end{equation*}
by Corollary \ref{cor:eoe}. Therefore, there is a strong
connection between the extended variational principle and the
Gibbs, de Finetti principle. We will make one more connection, by
constructing a joint ``Lagrangian''. The joint Lagrangian we
construct is the function
\begin{equation*}
  \mathcal{L}(\mu,\nu)\, =\, S_1(\mu) - \Phi(\nu) - \Phi^{(1)}(\nu,\mu-\nu)\, .
\end{equation*}
%(In principle there may be others.)

Since $g$ is concave and upper semicontinuous no matter what the
Hamiltonian, we see that $\mathcal{L}(\cdot,\nu)$ is concave and
upper semicontinuous for all $\nu$. Moreover, it is trivial to
check that
$$
  \max_{\mu}\, \mathcal{L}(\mu,\nu)\, =\, \tilde{g}(\nu)\, ,
$$
using Theorem \ref{thm:GVP}. Similarly, using convexity of $\Phi$
it is trivial to check that
\begin{equation*}
    \inf_{\nu}\, \mathcal{L}(\mu,\nu)\, =\, \min_{\nu}\,
\mathcal{L}(\mu,\nu)\, =\, g(\mu)\, .
\end{equation*}
The minimum is attained at $\mu = \nu$. This is by inequality
(\ref{convex:consequence}). In the concave case, the analogous
inequality proves that
\begin{equation*}
    \sup_{\nu}\, \mathcal{L}(\mu,\nu)\, =\, \max_{\nu}\,
\mathcal{L}(\mu,\nu)\, =\, g(\mu)\, .
\end{equation*}
The main purpose of this section is to prove Theorem
\ref{thm:extension}. For this purpose, we will use the following
generalization of von Neumann's minimax theorem. We refer to
\cite{Sion} for an elegant (rather {\em topological}) proof.

%\section{Extension of EVP in the Quadratic, Convex Case}
%
%The main result we want to point out is the following.
%
%\begin{prop}
%\label{prop:extension}
%Suppose that $\phi : \Omega^2 \to \R \cup \{+\infty\}$ is Borel measurable, bounded below,
%and $\Phi : \mathcal{M}_{+,1}(\Omega^2) \to \R \cup \{+\infty\}$ is convex.
%Then
%$$
%  p(\beta)\, =\, \inf_{\nu} \tilde{g}(\beta;\nu)\, ,
%$$
%where the infimum is taken over those $\nu$ such
%that $\Phi(\nu)<\infty$.
%\end{prop}

%\begin{rem}
%This allows one to treat the physically important problems
%of two-body repulsions, with a divergent form at the origin,
%as long as the repulsion is ``conditionally positive semidefinite''.
%See Section \ref{sec:examples}
%\end{rem}
%
%The tool which allows us to prove this is the following theorem.

\begin{thm}[Kneser, Fan Minimax Theorem]
\label{thm:KF} Let $\mathscr{M}$ be a compact, convex space and
let $\mathscr{N}$ be any convex space. Suppose that $\mathcal{L}$
is a function on $\mathscr{M}\times \mathscr{N}$ that is
concave-convex. If $\mathcal{L}$ is upper semicontinuous on
$\mathscr{M}$ for each $\nu \in \mathscr{N}$, then
\begin{equation*}
    \sup_{\mu \in \mathscr{M}}\, \inf_{\nu \in \mathscr{N}}\, \mathcal{L}(\mu,\nu)\,
    =\, \inf_{\nu \in \mathscr{N}}\, \sup_{\mu \in \mathscr{M}}\, \mathcal{L}(\mu,\nu)\, .
\end{equation*}
\end{thm}

\noindent {\bf Remark:} {\it The Kneser, Fan theorem generalizes
the ``von Neumann minimax theorem'' which is well-known as one of
the first mathematical results in game theory.}

\medskip
The definition of being concave-convex is that: for each $\nu \in
\mathscr{N}$ the function $\mathcal{L}(\cdot,\nu)$ should be
concave on $\mathscr{M}$, and for each $\mu \in \mathscr{M}$ the
function $\mathcal{L}(\mu,\cdot)$ should be convex on
$\mathscr{N}$. Note that in the $n=2$ case, we can write
\begin{equation*}
    \mathcal{L}(\mu,\nu)\, =\, S_1(\mu) + \Phi(\nu) -\Phi^{(1)}(\nu,\mu)\,
\end{equation*}
which is convex in $\nu$ as long as $\Phi$ is convex, because
$\Phi^{(1)}(\cdot,\mu)$ is linear. Therefore, in this case
$\mathcal{L}(\mu,\nu)$ is concave-convex. Among other things, this
means that $\tilde{g}$ is convex. One requirement for applying
Theorem \ref{thm:extension} is that the function $\mathcal{L}$ is
assumed to map into $\R$ (instead of $\R \cup \{\pm \infty\}$).
This is the reason that we stated Theorem \ref{thm:KF} in the
precise way we did.

\subsection{Proofs}

In order to prove Theorem \ref{thm:extension}, we will need more
information about the maximizer of $g$. Note that, since $g$ is
upper semicontinuous, and $\mathcal{M}^{+}_1(\Omega)$ is a compact
set, it does attain its maximum. If $\Phi$ is convex, then $g$ is
also strictly concave simply because
\begin{equation*}
    g(\mu)\, =\, S_1(\mu) - \Phi(\mu)\, ,
\end{equation*}
and $S_1$ is strictly concave. Therefore, the maximum is unique.
In order to state the following lemma, let $C_{\phi}$ be the
finite constant $C_{\phi}\, =\, \inf_{x \in \Omega} \phi(x)$. Let
$C_{\alpha}\, =\, \alpha^{\otimes n}(\phi)<\infty$. Note that
$C_{\alpha} \geq C_{\phi}$.
%Also note that $p_* \geq -C_{\alpha}$, and that
%$\Phi(\nu)>C_{\phi}$ for every $\nu \in
%\mathcal{M}^{+}_1(\Omega)$. Finally, $S_1(\mu)\leq 0$ for every $\mu$ as in
%Proposition \ref{prop:Properties:Entropy}.

\begin{lem} \label{lem:regularity}
Let $\mu_{*} \in \mathcal{M}^{+}_1(\Omega)$ be the maximizer of
$g$. Then $\mu \ll \alpha$ and
\begin{equation}
\label{EL1}
    \frac{d\mu_*}{d\alpha}(x)\, =\,
    \exp\left(C_*-\Phi^{(1)}(\mu_*,\delta_x)\right)
\end{equation}
for $\alpha$-a.e.\ $x \in \Omega$. Here, $C_*$ is a finite
constant related to $\mu_*$ and $p_*$ by
\begin{equation*}
    C_*\, =\, \Phi(\mu_*) - p_*\, =\, S_1(\mu_*) - 2 p_*\, .
\end{equation*}
In particular, one has the bounds
\begin{equation*}
    C_{\alpha} + C_{\phi}\, \leq\, C_*\, \leq\, 2
    C_{\alpha}\, ,
\end{equation*}
so that
\begin{equation*}
    \left\|\frac{d\mu_*}{d\alpha}\right\|_{\infty}\, \leq\,
    \exp\left(2[C_{\alpha} - C_{\phi}]\right)\, .
\end{equation*}
\end{lem}

\medskip
\noindent {\bf Proof.} Note that $g$ is finite on $\alpha$, so
that $g(\mu_*)>-\infty$. In particular, this means that
$S_1(\mu_*)>-\infty$. So $\mu_{*} \ll \alpha$. Suppose, in order
to reach a contradiction, that $\supp(\alpha)\setminus
\supp(\mu_{*}) \neq \emptyset$. Then there is a ball
$B=B(x;r)\subset\Omega$, $r>0$, such that $\mu_{*}(B)=0$ and
$\alpha(B)>0$. Let
\begin{equation*}
    \nu\, :=\, \alpha(B)^{-1}\, \chi_B\, \alpha\, ,
\end{equation*}
where $\chi_B$ is the indicator function of $B$. Let
\begin{equation*}
    \mu_\epsilon\, :=\, (1-\epsilon)\cdot \mu_* + \epsilon \cdot \nu\, .
\end{equation*}
A straightforward calculation shows that
\begin{equation*}
    \lim_{\epsilon\downarrow 0} \epsilon^{-1}\, [S_1(\mu_\epsilon)-S_1(\mu_{*})]\,
    =\, +\infty\, ,
\end{equation*}
whereas
\begin{equation*}
    \lim_{\epsilon \downarrow 0} \epsilon^{-1}\, [\Phi(\mu_\epsilon)-\Phi(\mu_*)]\,
    =\, \Phi^{(1)}(\mu_{*},\nu-\mu_*)
\end{equation*}
is a finite number. Hence there is an $\epsilon>0$ small enough so
that $ g(\mu_\epsilon)>g(\mu_*)$, contradicting the fact that
$\mu_*$ is a maximizer.

Now let $B=B(x_0;r)\subset\Omega$, for some $x_0 \in \supp(\mu_*)$
and $r>0$. Let
\begin{equation*}
    \nu\, :=\, \mu_*(B)^{-1}\, \chi_B\, \mu_*\, .
\end{equation*}
For $t \in \R$, let
\begin{equation*}
    \mu_{t}\, =\, (1-t)\cdot \mu_* + t\cdot \nu\, .
\end{equation*}
Note that for $-\mu_*(B)<t<1$, one has that $\mu_{t} \in
\mathcal{M}_{+,1}(\Omega)$. It is easy to see that the following
function is continuously differentiable,
\begin{equation*}
    \gamma(t)\,
    :=\, g(\mu_t)\,
    =\, S_1(\mu_t) - \Phi(\mu_t)\, .
\end{equation*}
Moreover, the derivative at $0$ is
\begin{equation*}
    \gamma'(0)\, =\, \int_{\Omega}
    \left[-\log\left(\frac{d\mu_*}{d\alpha}(x)\right)\right]\,
    d\nu(x) - \Phi^{(1)}(\mu_*,\nu) - S_1(\mu_*) +
    \Phi^{(1)}(\mu_*,\mu_*)\, .
\end{equation*}
By criticality, this must equal 0. So
\begin{equation*}
    \int_{\Omega}
    \log\left(\frac{d\mu_*}{d\alpha}(x)\right)\,
    d\nu(x) + \Phi^{(1)}(\mu_*,\nu)\, =\, C_*\, ,
\end{equation*}
where
\begin{equation*}
    C_*\, =\, \Phi^{(1)}(\mu_*,\mu_*)-S_1(\mu_*)\, =\, g(\mu_*) - \Phi(\mu_*)\, ,
\end{equation*}
is independent of $x$ and $r$. Note that since $g(\mu_*)=p_*$,
this gives the previous formulas for $C_*$. Note that
\begin{multline*}
    \int_{\Omega}
    \log\left(\frac{d\mu_*}{d\alpha}(x)\right)\,
    d\nu(x) + \Phi^{(1)}(\mu_*,\nu)\, -C_*\\ =\,
    \int_{\Omega}
    \left[\log\left(\frac{d\mu_*}{d\alpha}(x)\right) + \Phi^{(1)}(\mu_*,\delta_x) - C_*\right]\,
    d\nu(x)\, .
\end{multline*}
Since the total integral equals zero for all $\nu$, and $x_0$ and
$r$ are arbitrary, one concludes that
\begin{equation*}
    \log\left(\frac{d\mu_*}{d\alpha}(x)\right) + \Phi^{(1)}(\mu_*,\delta_x) -
    C_*\, =\, 0\, ,
\end{equation*}
for almost every $x \in \supp(\mu_*)$. But
$\supp(\mu_*)=\supp(\alpha)$. Exponentiating this equation yields
(\ref{EL1}). $\blacksquare$

\medskip
\noindent {\bf Proof of Theorem \ref{thm:extension}.} Observe
that, for any $0\leq C<\infty$, the subset
$\mathcal{M}^{+}_1(\Omega,\alpha,C)$ is compact and convex in
$\mathcal{M}^{+}_1(\Omega)$. Also,  $\mathcal{L}(\mu,\nu)$ is
well-defined and finite for all $\mu,\nu \in
\mathcal{M}^{+}_1(\Omega,\alpha,C)$. Part of this statement is
that $\Phi(\nu)$ and $\Phi^{(1)}(\nu,\mu)$ are finite. This is
tantamount to the first remark following the statement of Theorem
\ref{thm:extension}. The other fact is that $S_1(\mu)$ is finite,
because $S_1(\mu)\geq \psi(e^{C})>-\infty$. Therefore, the
hypotheses of Theorem \ref{thm:KF} are satisfied, so that
\begin{equation}
\label{KFapp}
    \sup_{\mu \in \mathscr{M}}\, \inf_{\nu \in \mathscr{N}}\, \mathcal{L}(\mu,\nu)\,
    =\, \inf_{\nu \in \mathscr{N}}\, \sup_{\mu \in \mathscr{M}}\, \mathcal{L}(\mu,\nu)\,
    ,
\end{equation}
when $\mathscr{M}=\mathscr{N}=\mathcal{M}^{+}_1(\Omega,\alpha,c)$.

By inequality (\ref{convex:consequence}), for any $\mu \in
\mathcal{M}^{+}_1(\Omega)$,
\begin{equation*}
    \inf_{\nu \in \mathcal{M}^{+}_1(\Omega)}\, \mathcal{L}(\mu,\nu)\, =\, g(\mu)\, .
\end{equation*}
Moreover, the minimum is attained at $\mu = \nu$. In particular,
if $\mu \in \mathcal{M}^{+}_1(\Omega,\alpha,C)$, then so is the
minimizer $\nu$. I.e.,
\begin{equation*}
    \inf_{\nu \in \mathscr{N}}\, \mathcal{L}(\mu,\nu)\, =\,
    g(\mu)\,
\end{equation*}
for all $\mu \in \mathscr{M}$. Therefore,
\begin{equation}
\label{KFapp2}
    \sup_{\mu \in \mathscr{M}}\, g(\mu)\,
    =\, \inf_{\nu \in \mathscr{N}}\, \sup_{\mu \in \mathscr{N}}\, \mathcal{L}(\mu,\nu)\,
    ,
\end{equation}
by (\ref{KFapp}).

By Theorem \ref{thm:GVP},
\begin{equation}
\label{GVPapp}
  \sup_{\mu \in \mathcal{M}^{+}_1(\Omega)}\, \mathcal{L}(\mu,\nu)\, =\, \tilde{g}(\nu)\, ,
\end{equation}
for any $\nu \in \mathcal{M}^{+}_1(\Omega)$, by viewing
$\Phi^{(1)}(\nu,\mu)$ as a ($\nu$-dependent) Hamiltonian
integrated against $\mu$. So, optimizing over the smaller set
gives the inequality
\begin{equation*}
  \sup_{\mu \in \mathscr{N}}\, \mathcal{L}(\mu,\nu)\, \leq\, \tilde{g}(\nu)\,
  .
\end{equation*}
Therefore,
\begin{equation} \label{KFapp3}
    \sup_{\mu \in \mathcal{M}^{+}_1(\Omega,\alpha,c)}\, g(\mu)\,
    \leq\, \inf_{\nu \in \mathcal{M}^{+}_1(\Omega,\alpha,C)}\, \tilde{g}(\nu)\,
    ,
\end{equation}
by (\ref{KFapp2}).

By Lemma \ref{lem:regularity}, the unrestricted optimizer of $g$,
over $\mathcal{M}^{+}_1(\Omega)$ is $\mu_*$ which is in
$\mathcal{M}^{+}_1(\Omega,\alpha,C)$ for every
$C>2(C_{\alpha}-C_{\phi})$. Moreover, $g(\mu_*)=p_*$. So, by
(\ref{KFapp3}),
\begin{equation*}
    p_*\, \leq\, \inf_{\nu \in
    \mathcal{M}^{+}_1(\Omega,\alpha,C)}\, \tilde{g}(\nu)\, ,
\end{equation*}
for every $C>2(C_{\alpha}-C_{\phi})$. In particular,
\begin{equation}
\label{ineq:ext1}
    \lim_{C \to \infty}\, \inf_{\nu \in
    \mathcal{M}^{+}_1(\Omega,\alpha,C)}\, \tilde{g}(\nu)\,
    \geq\, p_*\, .
\end{equation}
The proof will be completed by also establishing the opposite
inequality.

As noted, $\mu_*$ is in $\mathcal{M}^{+}_1(\Omega,\alpha,C)$ for
$C>2(C_{\alpha}-C_{\phi})$. Therefore,
\begin{equation}
\label{ineq:sddl}
    \lim_{C \to \infty}\, \inf_{\nu \in
    \mathcal{M}^{+}_1(\Omega,\alpha,C)}\, \tilde{g}(\nu)\,
    \leq\, \tilde{g}(\mu_*)\, .
\end{equation}
By (\ref{gtilde:def}),
\begin{equation*}
  \tilde{g}(\mu_*)\,
  =\, \Phi(\mu_*)
  + \log\, \int_{\Omega} \exp\left(-\Phi^{(1)}(\mu_*;\delta_x)\right)\, d\alpha(x)\, .
\end{equation*}
But, by equation (\ref{EL1}),
\begin{equation*}
    \int_{\Omega} \exp\left(-\Phi^{(1)}(\mu_*;\delta_x)\right)\, d\alpha(x)\,
    =\, \exp(-C_*)\, \int_{\Omega} \frac{d\mu_*}{d\alpha}(x) \, d\alpha(x)\,
    =\, \exp(-C_*)\, .
\end{equation*}
Therefore,
\begin{equation*}
  \tilde{g}(\mu_*)\,
  =\, \Phi(\mu_*) - C_*\, .
\end{equation*}
But also by Lemma \ref{lem:regularity}
\begin{equation*}
    \Phi(\mu_*) - C_*\, =\, p_*\, .
\end{equation*}
Therefore, combining with (\ref{ineq:sddl}),
\begin{equation*}
    \lim_{C \to \infty}\, \inf_{\nu \in
    \mathcal{M}^{+}_1(\Omega,\alpha,C)}\,
    \leq\, p_*\, ,
\end{equation*}
as needed. $\blacksquare$

\medskip
\noindent {\bf Remark:} {\it {\bf A posteriori} it is clear that
there is a saddle point for the Lagrangian $\mathcal{L}(\mu,\nu)$
at $\mu=\nu=\mu_*$. However since $\mathcal{L}$ may not be
strictly concave-convex, this may not be the only argminmax or
argmaxmin. (C.f., \cite{RockafellarWets}, Chapter 11, Sections I
and J, for the relevant notation from convex variational
analysis.) If one could establish that $\tilde{g}$ has an
optimizer which can be identified by the Euler-Lagrange equations,
then it must also be an optimizer for $g$ because the
Euler-Lagrange equations are the same. However, except in the case
that $\phi$ is bounded and continuous, it is not clear that this
is the case {\bf a priori}.}

%%%%%%%%%%%%%%%%%%%%%%%%%%%%%%%%%%%%%%%%%%%%%%%%%%%%%%%%%
%%%%%%%%%%%%%%%%%%%%%%%%%%%%%%%%%%%%%%%%%%%%%%%%%%%%%%%%%
%%%%%%%%%%%%%%%%%%%%%%%%%%%%%%%%%%%%%%%%%%%%%%%%%%%%%%%%%

\section{Example: The Negative Quadratic Kernel}
\label{sec:examples}

Let us consider $\Omega \subset \R^d$ compact, and $\phi(x,y) =
-\|x-y\|^2$. It is well-known that this defines a positive
semidefinite form $\overline{\Phi}^{(1)} : \mathcal{M}_0(\Omega)
\times \mathcal{M}_0(\Omega) \to \R$ by the map
$\overline{\Phi}^{(1)}(\mu,\nu)\, =\, 2(\mu\otimes \nu)(\phi)$,
where $\mathcal{M}_0(\Omega)$ is the set of all bounded-variation,
signed measures with total measure equal to 0. (In fact this is
the critical homogeneous potential with this property. C.f.,
Schoenberg \cite{Schoenberg}.) Therefore, $\Phi$ is convex. We
note that, for $\nu \in \mathcal{M}^{+}_1(\Omega)$
\begin{align*}
    \frac{1}{2} \Phi^{(1)}(\nu,\delta_x)\,
    &=\, - \int_{\Omega} \|x - y\|^2\, d\nu(y)\\
    &=\, - \|x\|^2 + 2\int_{\Omega} (x,y)\, d\nu(y) -
    \int_{\Omega} \|y\|^2\, d\nu(y)\\
    &=\, - \|x - \E[X]\|^2 - \Var(X)\, ,
\end{align*}
where, $X$ is a random variable which is $\nu$-distributed. Using
this, we also have
\begin{align*}
    \Phi(\nu)\, &=\, - \Var(X)  - \int_{\Omega} \|x -
    \E[X]\|^2\, d\nu(x)\\
    &=\, - 2 \Var(X)\, .
\end{align*}
Therefore,
\begin{align*}
    \tilde{g}(\nu)\,
    &=\, \Phi(\nu)
    + \log\, \int_{\Omega} \exp\left(-\Phi^{(1)}(\nu,\delta_x)\right)\, d\alpha(x)\\
    &=\, - 2 \Var(X)
    + \log\, \int_{\Omega} \exp\left(2 \Var(X)+2\|x-\E[X]\|^2\right)\, d\alpha(x)\\
    &=\, \log\, \int_{\Omega} \exp\left(2\|x-\E[X]\|^2\right)\, d\alpha(x)\, .
\end{align*}
In particular, this only depends on $\nu$ through $\E^{\nu}[X]$.
(We will write $\E^{\nu}[X]$ when we want to specify that $X$ is
$\nu$-distributed.) Given any $x_0 \in \Omega$, we can choose $\nu
= \delta_{x_0}$ so that there is at least one $\nu$ such that
$\E^{\nu}[X]=x_0$.  Therefore, the extended variational principle
tells us that
\begin{equation*}
    p_*\, =\, \min_{y \in \Omega}\, \log\, \int_{\Omega}
    \exp\left(2 \|x-y\|^2\right)\, d\alpha(x)\, .
\end{equation*}

This is obviously a convex optimization problem, where the convex
cost functional to be minimized is
\begin{equation*}
    \mathcal{C}(y)\, =\,  \log\, \int_{\Omega}
    \exp\left(2 \|x-y\|^2\right)\, d\alpha(x)\, .
\end{equation*}
Moreover, since $\Omega$ is compact and since the cost functional
is continuous, there does exist a unique solution. Notice that the
criticality condition is the implicit characterization:
\begin{equation*}
    y\, =\, \frac{\int_{\Omega} x\, e^{2 \|x-y\|^2}\,
    d\alpha(x)}{\int_{\Omega} e^{2 \|x-y\|^2}\, d\alpha(x)}\,
    .
\end{equation*}
This example contains mean-field Ising and Heisenberg {\em
antiferromagnets} as special cases. These are obtained by taking
$\Omega = \mathbb{S}^{d-1}$, the spheres in $\R^d$. The Ising case
is $d=1$ for which we have $\mathbb{S}^0 = \{-1,+1\}$. We can
include a one-body term, representing and external magnetic field,
by a special choices of the {\em a priori} measure. We can also
determine the Gibbs measure. It is equal to
$$
  \frac{d\rho_*}{d\alpha}(x)\, =\, Z^{-1} e^{2\|x-x_*\|^2}\, ,
$$
where $y=x_*$ solves the optimization problem above.

If we change $\phi$ to $-\phi$, we obtain the {\em ferromagnetic}
version of these mean-field models. However, the analogous cost
function becomes
\begin{equation*}
    \mathcal{C}(y)\, =\,  \log\, \int_{\Omega}
    \exp\left(-2 \|x-y\|^2\right)\, d\alpha(x)\, .
\end{equation*}
and we have $p_*\, =\, \max_{y \in \Omega} \mathcal{C}(y)$. Since
the cost function is not concave, there can be multiple optimizers
(depending on $\Omega$ and $\alpha$) which may be interpreted as
the existence of a {\em phase transition}.

\section*{Acknowledgements}
S.S.\ is most grateful to Michael Aizenman and Bob Sims,
especially to Michael Aizenman who invented the EVP. We also
benefitted from discussions with the following people: Aernout van
Enter, Bruno Nachtergaele, Vojkan Jaksic, Mark Fannes, Marco
Merkli and Ugur G\"ul. In particular, A.C.D.~van Enter and
B.~Nachtergaele alerted us to Lemma
\ref{lem:Monotone:Entropy:Density} as well as gave other help. The
research of E.K.\ was supported in part by FQRNT.

\end{document}